\documentclass[authoryear,preprint,12pt,]{elsarticle}

\usepackage{amssymb}
\usepackage{upgreek}
\usepackage{lineno}
\usepackage{threeparttable}
\usepackage{setspace}
\usepackage{sidecap}
\usepackage{longtable}
\usepackage{multirow}
\usepackage{booktabs}
\usepackage{caption}
\usepackage{wasysym}
\usepackage{rotating}

\journal{Geochimica et Cosmochimica Acta}

\begin{document}

\begin{frontmatter}



\title{Zinc isotope analyses of singularly small samples ($<$5 ng Zn): investigating chondrule-matrix complementarity in Leoville}


\author{Elishevah van Kooten}

\address{Universit\'{e} de Paris, Institut de Physique du Globe de Paris, 	CNRS UMR 7154, 1 rue Jussieu, 75238 Paris, France}

\author{Fr\'{e}d\'{e}ric Moynier}

\address{Universit\'{e} de Paris, Institut de Physique du Globe de Paris, 	CNRS UMR 7154, 1 rue Jussieu, 75238 Paris, France; Institut Universitaire de France, Paris, France}

\begin{abstract}
The potential complementarity between chondrules and matrix of chondrites, the Solar System's building blocks, is still a highly debated subject. Complementary superchrondritic compositions of chondrite matrices and subchondritic chondrules may point to formation of these components within the same reservoir or, alternatively, to mobilization of elements during secondary alteration on chondrite parent bodies. Zinc isotope fractionation through evaporation during chondrule formation may play an important role in identifying complementary relationships between chondrules and matrix and is additionally a mobile element during hydrothermal processes. In an effort to distinguish between primary Zn isotope fractionation during chondrule formation and secondary alteration, we here report the Zn isotope data of five chondrule cores, five corresponding igneous rims and two matrices of the relatively unaltered Leoville CV3.1 chondrite. The detail required for these analyses necessitated the development of an adjusted Zn isotope analyses protocol outlined in this study. This method allows for the measurement of 5 ng Zn fractions, for which we have analyzed the isotope composition with an external reproducibility of 120 ppm. We demonstrate that we measure primary Zn isotope signatures within the sampled fractions of Leoville, which show negative $\updelta$$^{66}$Zn values for the chondrule cores ($\updelta$$^{66}$Zn = --0.43$\pm$0.14 $\permil$), more positive values for the igneous rims ($\updelta$$^{66}$Zn = --0.01$\pm$0.30 $\permil$) and chondritic values for the matrix ($\updelta$$^{66}$Zn = 0.19$\pm$0.14 $\permil$). In combination with elemental compositions and petrology of these chondrite fractions, we argue that chondrule cores, igneous rims and matrix could have formed within the same reservoir in the protoplanetary disk. The required formation mechanism involves Zn isotope fractionation through sulfide loss during chondrule core formation and concurrent thermal processing of matrix material. Depleted olivine-bearing grains representing this processed matrix would have accreted to the depleted chondrule cores and subsequently reabsorbed material (including $^{66}$Zn-rich) from a complementary volatile-rich gas, thereby forming the igneous rims. This would have allowed the rims to move towards an isotopically chondritic composition, similar to the non-processed matrix in Leoville. We note that Zn isotope analyses of components in other chondrites (f.e., CM, CO, EC) are necessary to identify if this complementarity relationship is generic or unique for each chondrite group. The development of a Zn isotope protocol for singularly small samples is a step forward in that direction.

\end{abstract}

\begin{keyword}
Zn isotopes \sep CV chondrules \sep igneous rims 
\end{keyword}

\end{frontmatter}


\section{Introduction}
Chondrites represent the remnant building blocks of our Solar System and have been relatively unaltered since their formation approximately 4.56 billion years ago. Most chondrites mainly consist of chondrules, once molten silicate droplets formed by transient heating events in the protolanetary disk. They are cemented together by submicron-sized dust grains called matrix and further contain less abundant refractory inclusions (the first condensed solids, i.e., CAIs and AOAs, \citealp{Connelly2012}) and FeNi metal \citep{Krot2009,Scott2014}. Even though chondrites have never been molten and, hence, maintain their initial accretionary texture, most chondrites have been modified to some degree on their parent bodies by thermal metamorphism, aqueous alteration or both. These changes affect the chemical and isotope composition of bulk chondrites as well as of individual components by processes such as diffusion, fluid interaction and evaporation. In addition, chondrules may have experienced volatile-loss (or perhaps gain) and concurrent stable isotope fractionation during their formation and before their accretion into chondrite parent bodies \citep{Ebel2018}. To identify these different processes is paramount to decipher the pre-accretionary history of chondritic components from post-accretionary secondary alteration of asteroids.\\

Zinc is a moderately volatile element with a relatively low condensation temperature (T$_{c}$ = 726 K, \citealp{Lodders2003}) and is well-suited element to study volatility-related processes during the formation of the Solar System. Zinc isotope fractionation reported in bulk chondrites and achondrites has shed light on processes related to volatile-loss. Lunar basalts and un-brecciated eucrites show correlations between heavy isotope enrichment and volatile-depletion, consistent with kinetic fractionation during evaporation \citep{Paniello2012,Paniello2012a,Kato2015,Moynier2017}. Bulk carbonaceous chondrites on the other hand (with the exception of CH and CB chondrites), show light isotope enrichment with volatile depletion, a trend opposite to what is expected from these materials if they experienced evaporation \citep{Luck2005,Albarede2009,Pringle2017,Sossi2018}. Similarly, chondrules of the CV chondrite Allende are also enriched in light Zn isotopes, an observation attributed to sulfide loss during chondrule formation when heavy Zn isotopes preferentially partitioned into the sulfides \citep{Pringle2017}. \\

With the Zn isotope analyses of Allende chondrules, a first step has been taken into tracing chondrule formation and the link between Zn isotope variations of bulk carbonaceous chondrites and the relative abundance of their components. However, many new questions arise from this study. It is unclear, for example, how the high degree of hydrothermal alteration in Allende ($>$CV3.6; \citealp{Bonal2016}) relates to the Zn isotope variability in individual chondrules, since Zn is highly mobile in aqueous fluids. Moreover, chondrules from other chondrite groups might not show the same variability and magnitude of Zn isotope fractionation and it is unknown how this would relate to the bulk isotope signature of their parent bodies and the supposed complementarity between chondrules and matrix. Furthermore, multiple melting events of chondrules could have chemically and isotopically affected any initial relationships between chondritic components. These numerous issues require a detailed investigation of chondrule and matrix fractions, which is presently highly limited by the size of these constituents. Zinc isotope analyses usually rely on hundreds of ng Zn to reach an external reproducibility (2SD) of 40 ppm for $^{66}$Zn/$^{64}$Zn and 50 ppm for $^{68}$Zn/$^{64}$Zn \citep{Chen2013}. Recent analyses of Zn isotope ratios in seawater have reduced the amount of Zn to 6-7 ng with an external reproducibility of 200 ppm \citep{Conway2013}. However, the seawater matrix is completely different from solid rock materials and the use of double-spike in this method prevents the determination of mass-dependent relationships between $\updelta$$^{66}$Zn and $\updelta$$^{68}$Zn or the possible determination of non-mass dependent isotopic variations. \citet{Pringle2017} recently reported chondrule Zn isotope data with total Zn abundances between 30 and 500 ng, but without measurements of correspondingly low standards to test the accuracy and precision of these data. Here, we report a fine-tuned method to measure Zn isotopes in unprecedentedly small samples, with Zn abundances of 2-5 ng and with an external reproducibility of 120 ppm (5 ng Zn) for $^{66}$Zn/$^{64}$Zn for a range of terrestrial standards. We use this method in a proof of concept to demonstrate Zn isotope variations between individual chondrules of the relatively unaltered CV3.1 \citep{Bonal2016} chondrite Leoville, as well as relationships between core, igneous rim and surrounding matrix of these chondrules.

\section{Materials and methods}

\subsection{Materials}
We have sampled multiple terrestrial standards with varying bulk compositions and with known Zn isotope signatures. Per standard, five samples were weighed (50-200 $\upmu$g per sample) containing approximately 5-10 ng Zn each, from bulk powders of BCR-2 (USGS basalt), AGV2 (USGS andesite) and CP1 (metasediment from the Queyras Schiste lustrés Complex, \citealp{Inglis2017}). In addition, we have selected larger aliquots of BHVO-2 (USGS basalt) and CV chondrite NWA 12523 (5 mg), with totals of $\sim$600 ng Zn each. We have, furthermore, sampled five chondrule cores, five igneous rims and two surrounding matrices of the Leoville CV3.1 chondrite. This CV chondrite was selected based on its low degree of alteration relative to other CV chondrites \citep{Bonal2016}. Although Kaba has experienced a similarly low degree of thermal metamorphism, it has undergone pervasive aqueous alteration \citep{Keller1990}. The Leoville fractions were extracted by New Wave microdrill with tungsten carbide drill bits at the Institute de Physique du Globe de Paris (IPGP) and thereafter transferred to clean Savillex beakers following methods used in \citet{vankooten2017a}. The chondrule sizes ranged between 0.5 and 2 mm and care was taken not to drill too deep ($<$200 $\upmu$m) to avoid contamination from surrounding materials. All drillspots were carefully examined under a plain-light microscope and suspicious spots were discarded. We have drilled approximately three cylindrical holes (100 $\upmu$m diameter, 200 $\upmu$m depth) per sample, yielding 2-5 ng of Zn.\\

\subsection{Elemental composition}
\label{methods_composition}
To determine the elemental composition of the Leoville chondrule and matrix samples, we have analyzed major and minor element compositions by Agilent 7900 ICP-Q-MS at IPGP. Elemental compositions were analyzed from sample solutions taken from the 1.5M HBr cut acquired from anion column chromatography (see section \ref{methods_column}). Since sample digestion with HF acid leads to complete loss of Si, it was not possible to determine the total weight and corresponding weight percentages of elements in the samples. Hence, we present elemental compositions as ratios normalized to Mg. The terrestrial standards AGV2 and BCR-2 were measured alongside Leoville chondrule fractions to determine potential elemental fractionation on the columns. The percentual uncertainty was calculated by comparing these standard analyses to literature values. Besides bulk compositions of the sampled chondrule and matrix fractions, we have also acquired elemental maps using the Zeiss EVO MA10 scanning electron microscope (SEM) at IPGP.

\subsection{Column chromatography}
\label{methods_column}
We have developed a protocol for Zn isotope analyses, adapted from the method used by \citet{Moynier2006} to accommodate for the small sample sizes used in this work ($<$10 ng of Zn). The major limitations of measuring such small samples stem from 1) Zn contamination from acids used during purification, 2) from resin organics and elemental impurities that cause matrix effects on the mass spectrometer and 3) from low signal/noise ratios during isotope analyses. Hence, our adjusted purification protocol is aimed at reducing these factors to insignificant levels. The Zn blank produced in previous digestion and column chromatography methods for $>$1 $\upmu$g Zn aliquots is typically 5-20 ng \citep{Arnold2010,Chen2013,Moynier2015}, comprising 100-400 \% of a 5 ng sample. It is unclear what the major contributing factors to the blank are in previous studies, thus, we have carefully analyzed the Zn abundance in all acids used during our procedure (Table \ref{tab:blanks}). First, we have compared the Zn blank during sample digestion in 1 ml 3:1 concentrated HNO$_{3}$/HF mixtures, with and without the use of Parr bombs. The digestion with four different Parr bombs included 24 hours at 150$^{\circ}$C and 48 hours at 210$^{\circ}$C inside a furnace. The use of Parr bombs contributes $<$0.6 ng to the Zn blank, whereas digestion of samples excluding the Parr bombs adds insignificant amounts of Zn to the samples. Leoville samples have been digested in Parr bombs to ensure full assimilation of Zn-containing refractory spinels. Subsequently, the samples were dried down and left for 24 hours on a hotplate at 140$^{\circ}$C in 1 ml aqua regia, which contained negligible Zn blanks. After drying down the samples, they were taken up in 100 $\upmu$l 1.5M HBr and were left to flux on a hotplate for 1 hour before loading the samples onto the columns. 
\\

\begin{figure}[]
	\centering
	\includegraphics[width=0.9\textwidth]{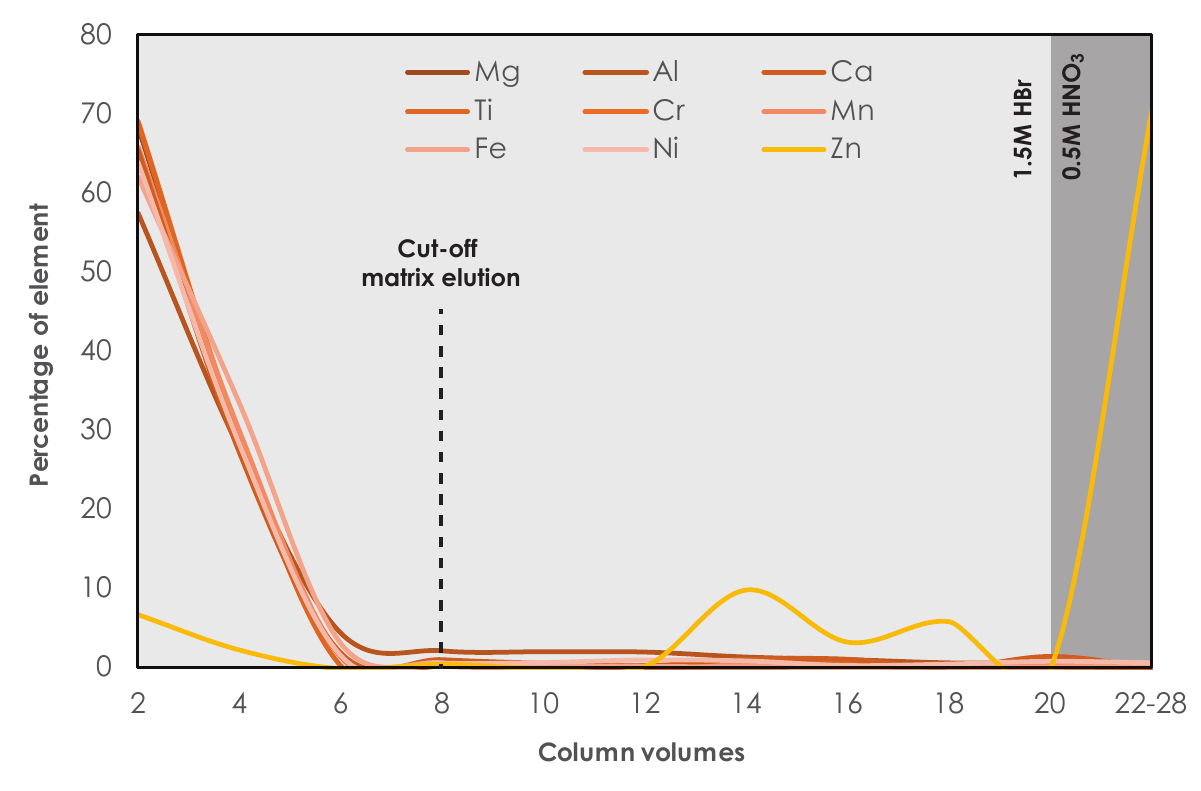}
	\caption{Elution profiles of major elements from a BHVO-2 standard. One column volume reflects 100 $\upmu$l of acid. After eight column volumes all matrix elements are eluted with 1.5M HBr. We note that Zn peaks between 12 and 19 column volumes are analytical artifacts due to a low signal/noise ratio. We have repeated these elution tests using BHVO-2 and CV chondrite aliquots spiked with 100 ng Zn. These tests resulted in reproducible yields $>$99.9 \%. 
		\label{fig:FigZnelutionprofile1}}
\end{figure}

\begin{figure}[]
	\centering
	\includegraphics[width=0.8\textwidth]{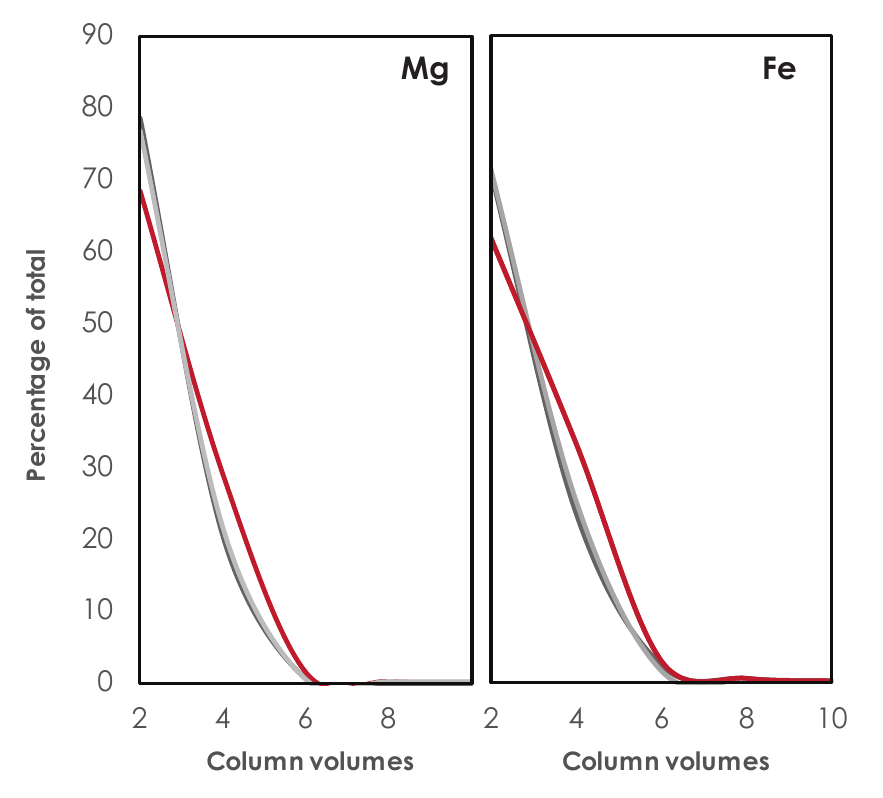}
	\caption{Elution profiles of Mg and Fe in 1.5M HBr for BHVO-2 (grey curves) and CV chondrite NWA 12523 (red curves). Repeat experiments show consistent elution profiles and purification of matrix elements after eight column volumes, independent of matrix composition. 
		\label{fig:FigZnelutionprofileMgFe}}
\end{figure}

\begin{figure}[]
	\centering
	\includegraphics[width=0.9\textwidth]{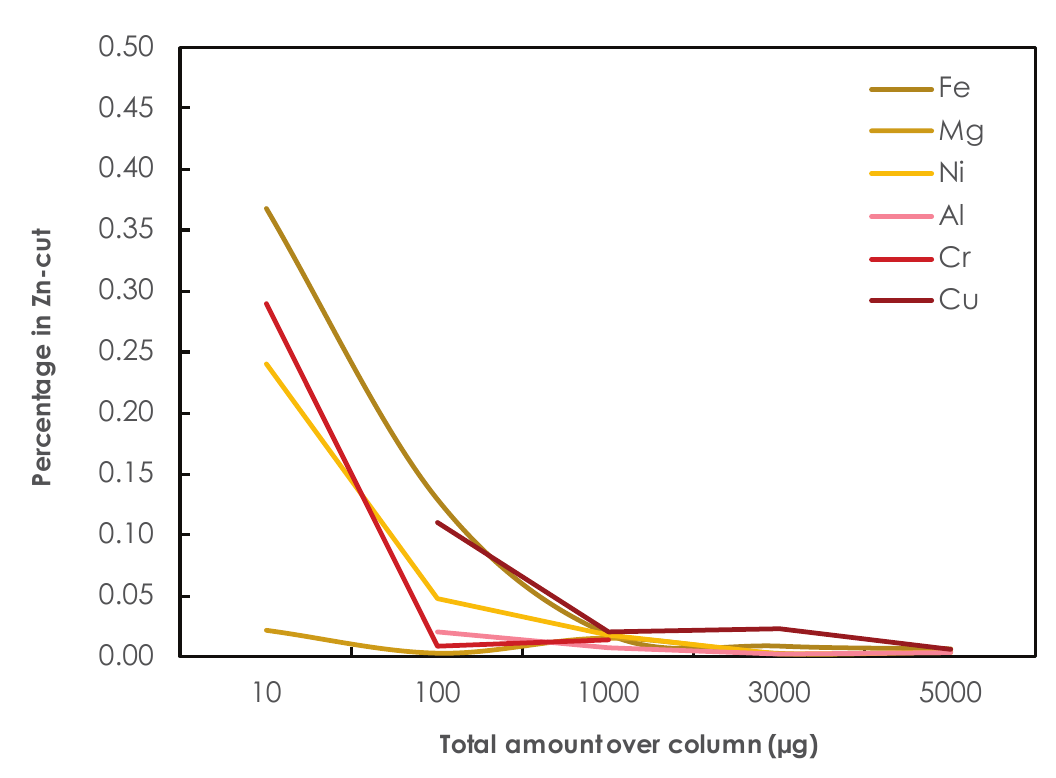}
	\caption{Analyses of several major and minor elements in the eluted Zn-cut from CV chondrite NWA 12523 after purification for various sample sizes, showing purification for each element $>$99.5 \%. Aliquots $>$100 $\upmu$g generally result in a cleaner Zn-cut. $>$99.9 \%. Sample sizes aimed at in this study ($\sim$50 $\upmu$g) have a relatively large abundance of matrix elements in the Zn-cut after a single column pass (Fe $\approx$30 ng and Ni $\approx$0.8 ng on 5 ng of Zn) and require a second column pass for acceptable Zn purification (Fe $\approx$0.08 ng and Ni $\approx$0.002 ng on 5 ng of Zn). 
		\label{fig:FigZncutcontamination}}
\end{figure}

Our Zn purification protocol employs an anion AG-1 X8 (200-400 mesh) exchange resin in teflon 4:1 shrinkage tubes, of which the resin bed is 2 mm $\times$ 2.5 mm, with a resin volume of 100 $\upmu$l. The columns are cleaned by alternating 7M HNO$_{3}$ with MQ-water three times, after which they are conditioned with 400 $\upmu$l 1.5M HBr. \citet{Moynier2015} use similar columns during a second and third purification step with a total acid volume of 11 ml per column (6 ml 1.5M HBr and 5 ml 0.5M HNO$_{3}$). We have carried out repeated elution tests using samples with varying compositions (e.g., BHVO-2 and CV chondrite NWA 12523) to reduce the amount of acids and, thus, the Zn contribution from these acids (Fig. \ref{fig:FigZnelutionprofile1} and \ref{fig:FigZnelutionprofileMgFe}). We show consistently that 800 $\upmu$l of 1.5M HBr is needed to elute $>$99.5 \% of the sample matrix (Fig. \ref{fig:FigZnelutionprofile1} and \ref{fig:FigZnelutionprofileMgFe}). Figure \ref{fig:FigZncutcontamination} shows that the size of the sample load over the column affects the matrix contribution to the Zn-cut. Samples smaller than $\sim$100 $\upmu$g have an increasingly contaminated Zn-cut by siderophile elements (e.g., Fe, Ni, Cr). Zinc is eluted in 800 $\upmu$l 0.5M HNO$_{3}$, with Zn yields $>$99.9 \%. This step is repeated once more to remove trace amounts of sample matrix. Over a period of four months we have analyzed five procedural blanks reflecting the Zn contamination of the column chromatography procedure. We find that the total column blank is non-reproducible, but in all cases $<$0.35 ng (on average $<$0.2 ng), which represents $<$7 \% of a 5 ng Zn sample. Our HBr and HNO$_{3}$ acid blanks are typically lower than the procedural blank ($<$0.01 ng), suggesting that the contamination is derived from random Zn impurities within the anion resin related to the initial manufacturing of the resin \citep{Shiel2009}, or perhaps from laboratory wear such as gloves \citep{Garcon2017}. Hence, the Zn procedural blank would only be lowered by use of a different resin, an even smaller column size or not at all. Nevertheless, for our purpose, the total Zn blank is surmountable for sample sizes of 5-10 ng Zn, since for observed natural Zn isotope fractionation ranges (i.e., $\sim$2 $\permil$, \citealp{Moynier2017}), the added error from blank contamination falls within the external reproducibility of our measurements (see section \ref{limitations}). We note that this purification procedure is suited for total sample sizes $<$5 mg. Loading sample solutions onto 100 $\upmu$l columns above this level resulted in a stagnation of acid elution.

\subsection{Neptune Plus MC-ICPMS}
Zinc isotope ratios were measured using a Thermo Scientific Neptune Plus Multi-Collector Inductively-Coupled-Plasma Mass-Spectrometer (MC-ICP-MS) at IPGP. The analytical setup is identical to \citet{Moynier2015}, barring the use of an APEX IF introduction system instead of a spray chamber. This allowed for a 3-4 times higher Zn signal. With one block of 30 measurements and an integration time of 8.389 seconds, 5 ng Zn aliquots in 500 $\upmu$l 0.1M HNO$_{3}$ (10 ppb) solutions could typically be measured at $\sim$0.7-1.0 V of $^{64}$Zn. Blank solutions were analyzed before and after measurements and generally contained $\sim$ 3-5 mV of $^{64}$Zn. At the beginning of each session (three sessions in total), the BHVO-2 standard (with a total of 600 ng Zn) was measured three times in aliquots of 5 ng Zn and bracketed by a JMC-Lyon standard. This was done to provide an indication of the potential accuracy and reproducibility of such small samples. All data are represented in the delta notation as permil deviations from the JMC-Lyon standard:

\begin{center}
	\[
	\updelta^\textrm{x}\textrm{Zn} =\left [ \frac{(^\textrm{x}\textrm{Zn}/^{64}\textrm{Zn})_\textrm{\scriptsize sample}}{(^\textrm{x}\textrm{Zn}/^{64}\textrm{Zn})_\textrm{\scriptsize JMC-Lyon}} -1 \right] \times 10^{3} 
	\]
\end{center}

\section{Results}
\subsection{Terrestrial standards}
\begin{figure}[]
	\centering
	\includegraphics[width=0.9\textwidth]{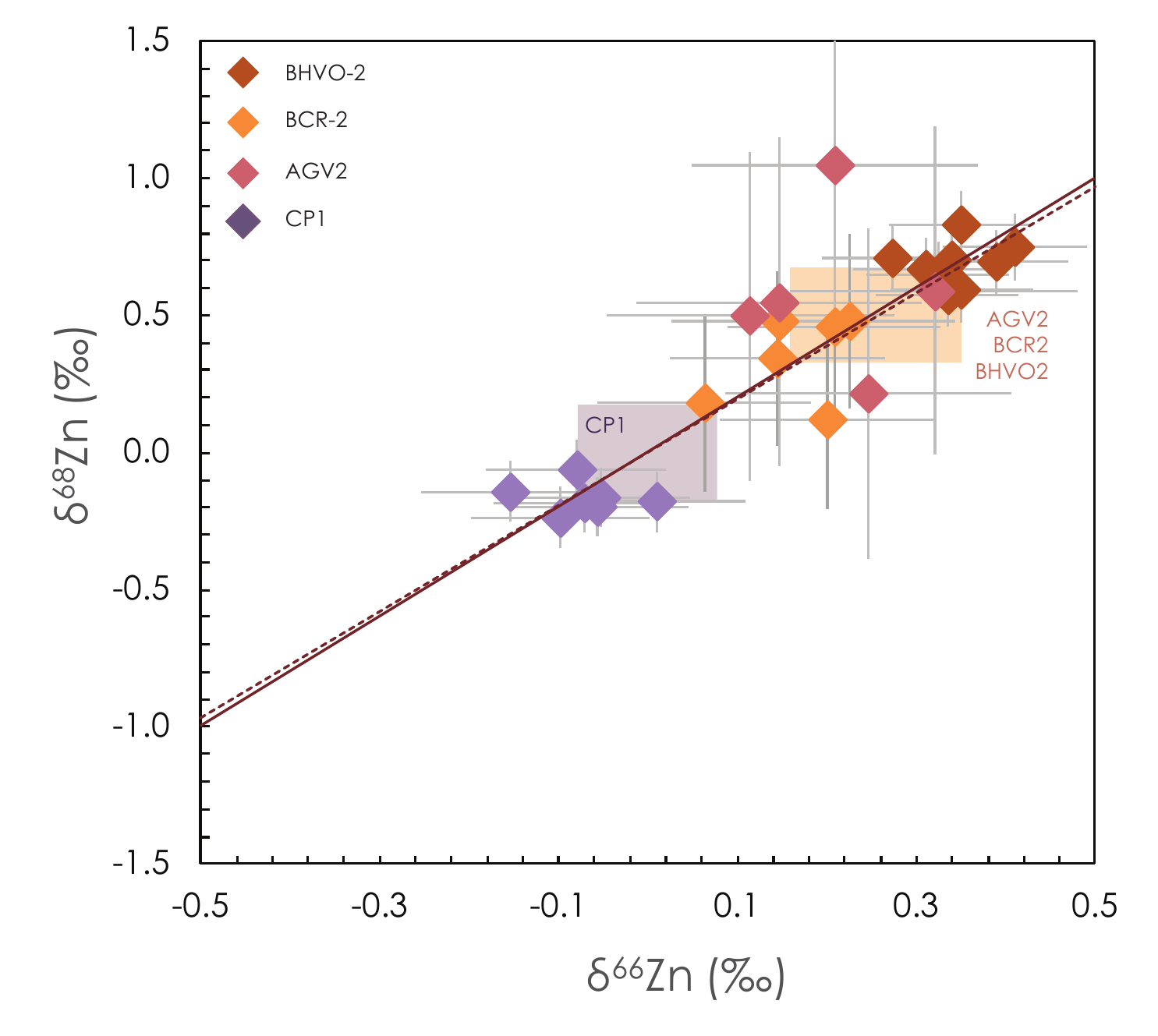}
	\caption{Zn isotope data showing $\updelta$$^{66}$Zn and $\updelta$$^{68}$Zn values for terrestrial standards and samples. Errors are 2SD from average mean (Table \ref{tab:standardsZn}). The shaded areas are (recommended) literature values for CP1 (purple, \citealp{Inglis2017}) and BHVO-2, BCR-2 and AGV2 (orange, \citealp{Moynier2017}). We note that literature only reports the $\updelta$$^{66}$Zn values. The solid and dashed lines represent kinetic and equilibrium fractionation laws, respectively.
		\label{fig:FigZnstandards}}
\end{figure}

Nine repeat analyses of the BHVO-2 standard with a total Zn abundance of 600 ng were carried out using fractions of 5 ng Zn per measurement. The resulting average $\updelta$$^{66}$Zn and $\updelta$$^{68}$Zn values were 0.34$\pm$0.08 $\permil$ (2SD) and 0.69$\pm$0.16 $\permil$ (2SD), respectively (Table \ref{tab:standardsZn}). The $\updelta$$^{66}$Zn value is in agreement with accepted literature values \citep{Moynier2017} of 0.29$\pm$0.09 $\permil$ (2SD) and $\updelta$$^{66}$Zn and $\updelta$$^{68}$Zn values define a mass-dependent relationship (Fig. \ref{fig:FigZnstandards}). The 2SD error of our measurements (80 ppm for $\updelta$$^{66}$Zn) is similar to that of repeat cross-laboratory analyses of BHVO-2 ($>$1 $\upmu$g Zn per measurement) and give us the expected lower limit external reproducibility of our analyses. Five Zn isotope analyses of BCR-2 from separate digestions of 5-10 ng Zn yield average $\updelta$$^{66}$Zn and $\updelta$$^{68}$Zn values of 0.16$\pm$0.12 $\permil$ (2SD) and 0.34$\pm$0.32 $\permil$ (2SD), respectively. A similar average $\updelta$$^{66}$Zn value is achieved for repeat measurements of AGV2 ($\updelta$$^{66}$Zn = 0.21$\pm$0.16 $\permil$, $\updelta$$^{68}$Zn = 0.58$\pm$0.60 $\permil$), although with a larger error on $\updelta$$^{66}$Zn and $\updelta$$^{68}$Zn. Both geological standards are within error of reported average literature values of $\updelta$$^{66}$Zn (0.25$\pm$0.08 $\permil$ for BCR-2, 0.29$\pm$0.06 $\permil$ for AGV2; \citealp{Moynier2017}). Our analyses of CP1 metasediments show that the obtained Zn isotope data ($\updelta$$^{66}$Zn = --0.07$\pm$0.10 $\permil$, $\updelta$$^{68}$Zn = --0.17$\pm$0.11 $\permil$) are also in agreement with previously reported data ($\updelta$$^{66}$Zn = 0.00$\pm$0.08 $\permil$, \citealp{Inglis2017}). All data lie on the mass-dependent correlation line (Fig. \ref{fig:FigZnstandards}). Even though all Zn isotope standards are in good agreement with reported values, we note that a small negative offset ($<$0.08 $\permil$ on $\updelta$$^{66}$Zn) may exist between our dataset and reported values. However, this potential offset is unresolvable, within error of the external reproducibility of our dataset, which we estimate to be 120 ppm on $\updelta$$^{66}$Zn. Although the uncertainty of AGV2 is higher than this estimation, this is likely the result of matrix element contamination to the Zn-cut (see section \ref{limitations}).

\subsection{Leoville chondrule cores, igneous rims and matrix}
\subsubsection{Petrology}
\begin{figure}[]
	\centering
	\includegraphics[width=0.6\textwidth]{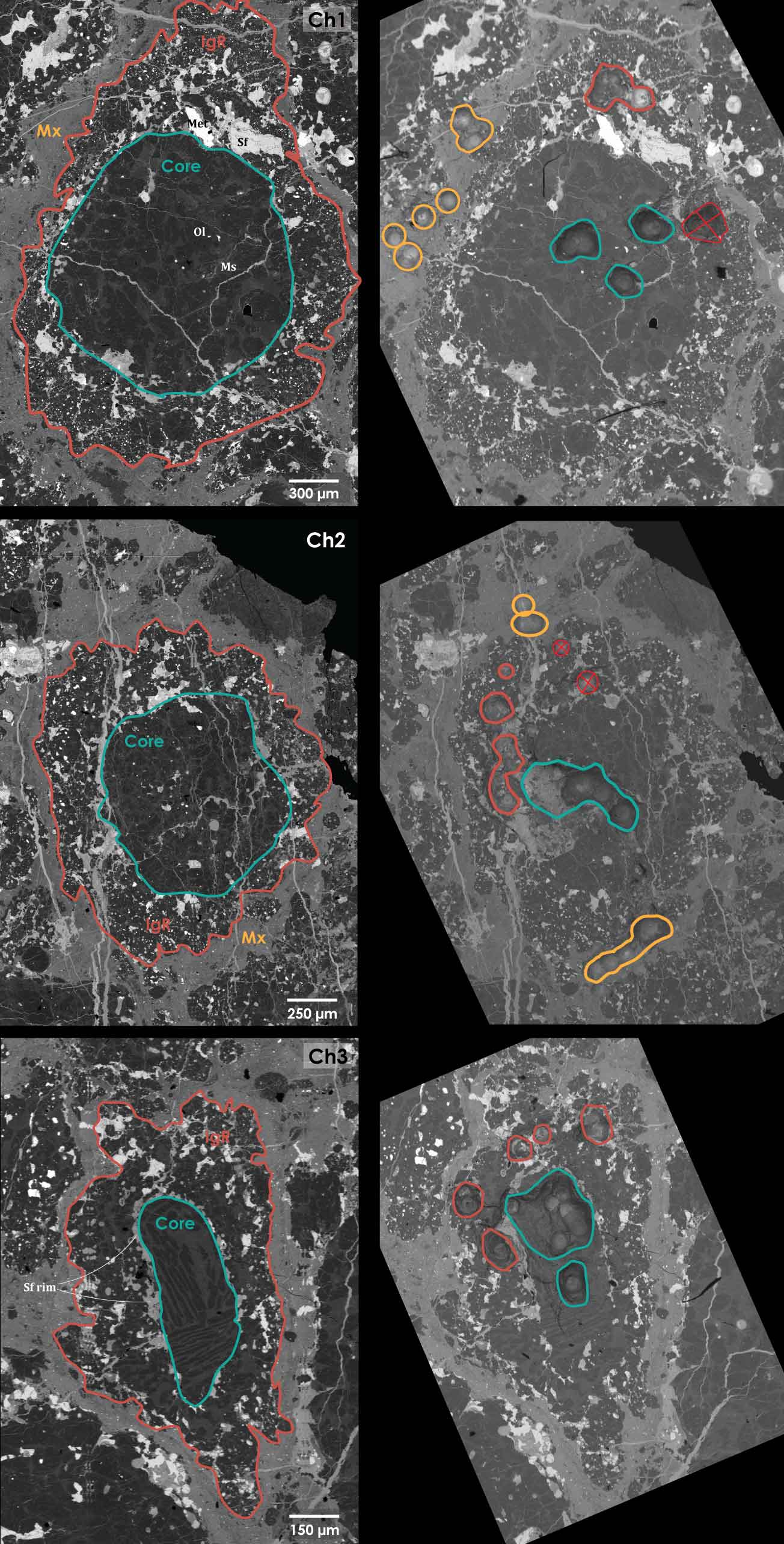}
	\caption{\textit{Left:} BSE images for Leoville chondrules in which their cores, igneous rims (IgR) and matrices (Mx) are highlighted. \textit{Met} = metal, \textit{Sf} = sulfide, \textit{Ol} = olivine, \textit{Ms} = mesostasis. \textit{Right:} BSE images of the chondrules after microdrilling. The drill locations for cores, rims and matrices are highlighted. Discarded drill holes are reflected by red coded areas.
		\label{fig:FigBSEchondrules1}}
\end{figure}

\begin{figure}[]
	\centering
	\includegraphics[width=0.9\textwidth]{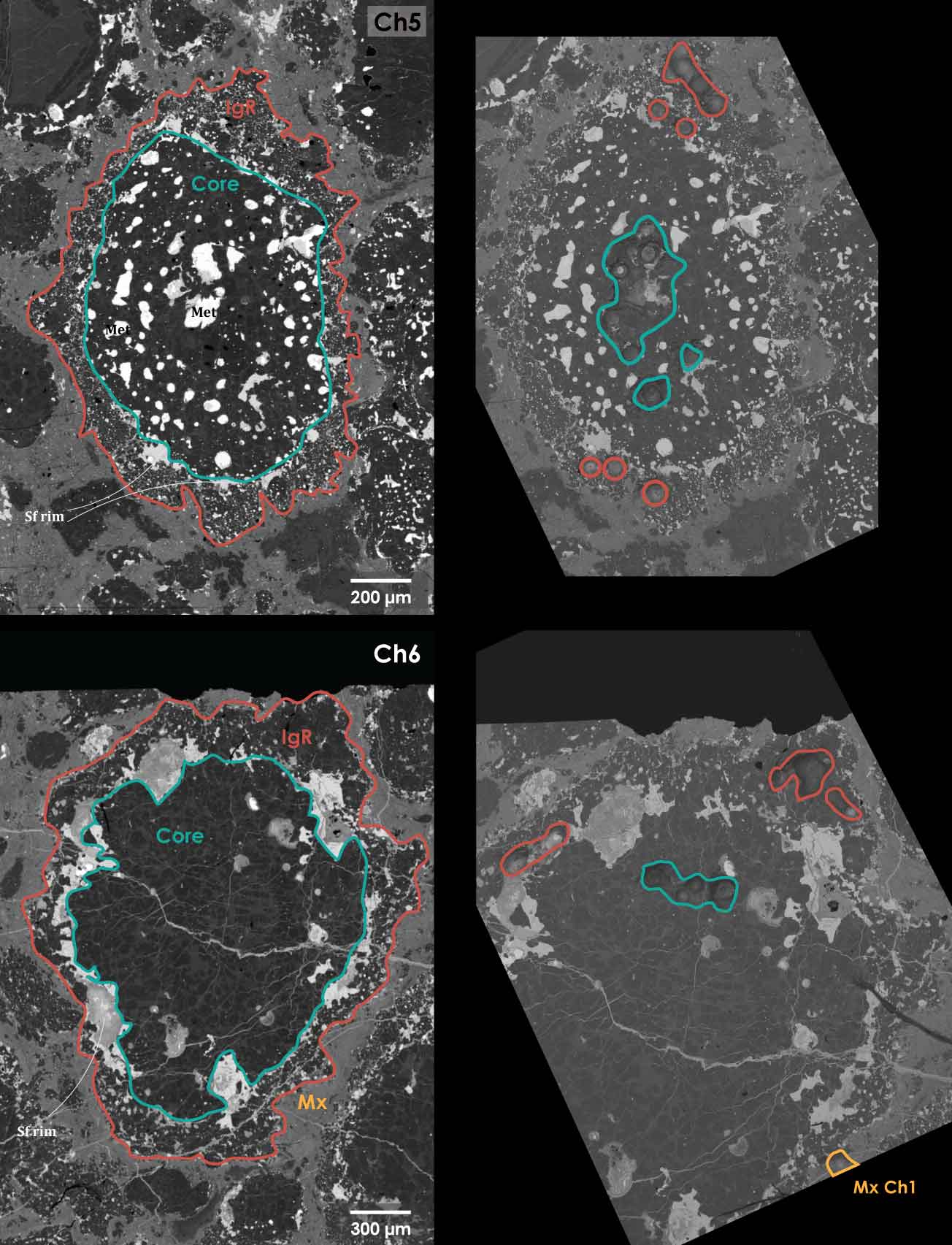}
	\caption{\textit{Left:} BSE images for Leoville chondrules in which their cores, igneous rims (IgR) and matrices (Mx) are highlighted. \textit{Met} = metal, \textit{Sf} = sulfide, \textit{Ol} = olivine, \textit{Ms} = mesostasis. \textit{Right:} BSE images of the chondrules after microdrilling. The drill locations for cores, rims and matrices are highlighted. Discarded drill holes are reflected by red coded areas.
		\label{fig:FigBSEchondrules2}}
\end{figure}

\begin{sidewaysfigure}[]
	\centering
	\includegraphics[width=1.1\textwidth]{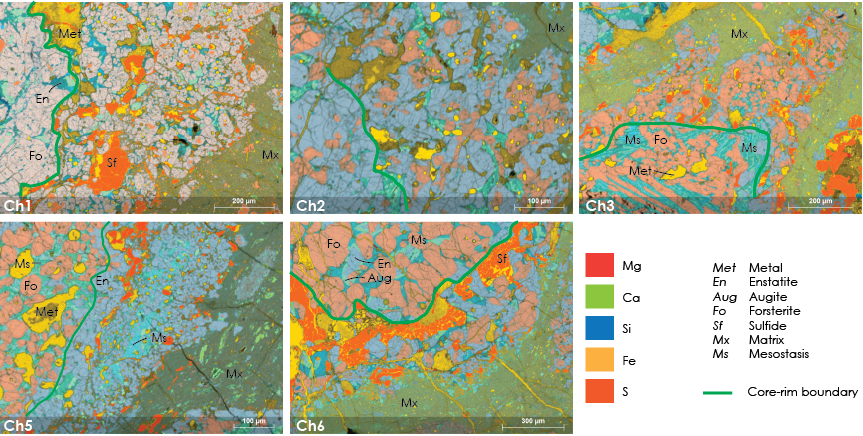}
	\caption{Elemental maps of chondrule areas showing Leoville cores, igneous rims and surrounding matrix. 
		\label{fig:MapsRims}}
\end{sidewaysfigure}

A total of five Leoville type-I (FeO-poor) chondrules were selected for Zn isotope analyses (Fig. \ref{fig:FigBSEchondrules1}, \ref{fig:FigBSEchondrules2} and \ref{fig:MapsRims}). Four out of five chondrule cores consist of porphyritic subhedral/euhedral forsterite inside a glassy mesostasis. Subhedral/euhedral enstatite has grown between olivine crystals, zoned by Ca-rich pyroxene. One chondrule core contains barred olivine with interspersed enstatite needles (Ch3). Most chondrule cores contain minor FeNi metal or troilite inclusions ($<$5 vol.\%), except for Ch5, which contains a large proportion of metal ($\sim$30 vol.\%). The cores are surrounded by sometimes thin and sharp (Ch3 and Ch5), but mostly irregular rings of FeNi metal and/or troilite aggregates (Ch6). In Ch1 and Ch2 the distinction between metal/sulfide rim and outer igneous rim cannot be made as the metal/sulfide is interspersed between the silicates. The igneous rims of the Leoville chondrules are more fine-grained than their cores. The rims consist of small 10-100 $\upmu$m anhedral to euhedral forsterite phenocrysts ($\sim$ 300 $\upmu$m in Ch6), which are to variable extent overgrown by low-FeO pyroxene (Ch1 $>$ Ch3 $>$ Ch2 $\approx$ Ch6 $>$ Ch5). The forsterite grains that are overgrown are typically anhedral and are indistinguishable in composition to the core olivines (i.e., low FeO content). The rim mineralogy is complemented by $<$100 $\upmu$m rounded metal and troilite grains, which appear identical in composition to the metal/sulfide rims between the cores and igneous rims. The rims typically contain more metal and especially troilite relative to the cores. These rim minerals are set in a glassy mesostasis that is similar in composition to that of the core, with the exception of Ch5, where the rim mesostasis is relatively Al-poor and Na-rich. The average thickness of the igneous rims is fairly constant between chondrules (200-300 $\upmu$m), even though the chondrule core size can vary significantly (0.2 - 2 mm). Within a rim, however, the thickness can vary considerably by a factor 5. Some chondrules also have surrounding fine-grained dust rims that show similar variations in thickness and these rims appear to complement the thickness of the igneous rims (Ch2 and Ch3). Hence, where the igneous rims are relatively thin, the dust rims are thicker. The boundaries between igneous and dust rims are generally irregular. Fractions of igneous rim can be found as inclusions in the dust rim (Ch3). The matrix locations from Ch1 and Ch2 are sampled close to the chondrule boundaries, although we could not distinguish between intra-chondrule matrix and dust rims. In most cases the dust rims of Leoville chondrules probably overlap with each other. In regions where clear dust rims are distinguished, a very narrow strip of intra-chondrule matrix seperates the two rims. Hence, in regions where chondrules lie closer together, the dust rims are directly connected to each other. The dust rims are generally more FeO-rich and fine-grained compared to the intra-chondrule matrix (Fig. \ref{fig:FigBSEchondrules2}, Ch3). The latter typically contains larger (mostly elongated) Fe sulfide and calcite grains. 
 
\subsubsection{Composition}
\label{composition}
Major and minor element compositions of Leoville chondrule fractions can be divided according to their 50\% condensation temperatures, where Ca, Al, and Ti are the most refractory elements and Co$>$Fe$>$Ni$>$Cr$>$Mn$>$Na are increasingly volatile \citep{Lodders2003}. These divisions show similar elemental patterns between cores, rims and matrices (Fig. \ref{fig:Majorelements}, Table \ref{tab:majorel}). For example, Ca/Mg, Al/Mg and Ti/Mg ratios are on average higher in the chondrules cores relative to the igneous rims, whereas they show a smaller variability and are near-chondritic (chondritic being the CI ratio in Fig. \ref{fig:Majorelements}) between rims and matrix. The cores have typically sub-chondritic Cr/Mg and Ni/Mg ratios, whereas the igneous rims are chondritic or super-chondritic. The Mn/Mg ratios of the cores are also sub-chondritic, whereas the igneous rims are close in composition to bulk CV chondrites and the matrix super-chondritic. The cores and rims have a larger spread in Na/Mg than in Mn/Mg ratios and both are close in composition to bulk CV chondrites. We note that for these volatile elements the bulk CV composition is sub-chondritic. The matrix Na/Mg ratios are sub-chondritic but higher than the cores and rims. Finally, the Fe/Mg ratios are sub-chondritic for the cores, chondritic for the rims and super-chondritic for the matrix. Overall, the igneous rims are very close in composition to bulk CV chondrites for a range of elements with different condensation temperatures, whereas the cores are super-chondritic for refractory elements and sub-chondritic for more volatile elements (with the exception of Na/Mg). The compositions of the rims are similar to those of the cores, except for Cr/Mg, Ni/Mg and Fe/Mg ratios, which can be linked to a higher abundance of metal and sulfide in the igneous rims. The matrix is generally chondritic or super-chondritic for a range of elemental compositions.

\subsubsection{Zinc isotope analyses}
\label{Results_Znisotopes}
\begin{figure}[]
	\centering
	\includegraphics[width=1.0\textwidth]{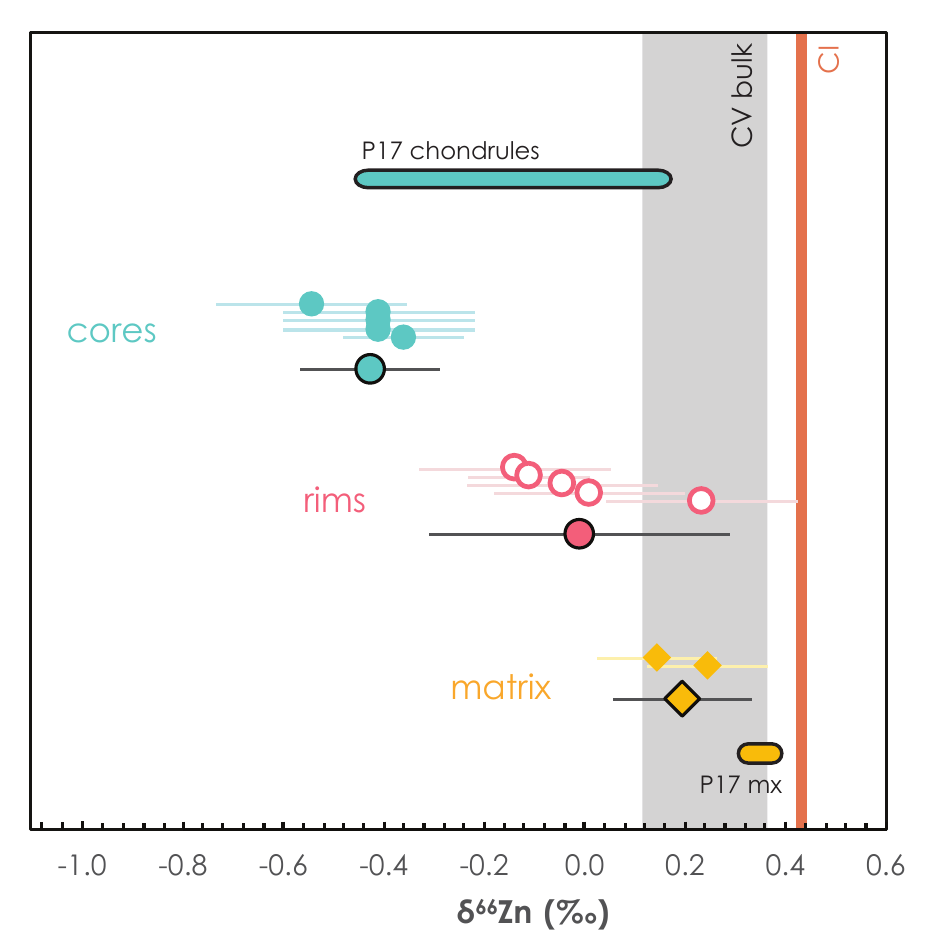}
	\caption{$\updelta$$^{66}$Zn values for Leoville cores, igneous rims and matrix, as well as their average compositions (black-rimmed symbols). Data for chondrules (blue) and matrix (yellow) from \citet{Pringle2017} (P17) is also shown. 
		\label{fig:ChondruleZnisotopes}}
\end{figure}

We report on the Zn isotope analyses of five chondrule cores, five igneous rims surrounding these cores and two matrix samples in the near vicinity of Ch1 and Ch2 (Fig. \ref{fig:ChondruleZnisotopes}, Table \ref{tab:LeovilleZn}). During the collection of the samples by microdrill it was difficult to estimate the total amount of Zn in each sample, since the Zn concentration in chondrules is variable and we could not accurately measure the total weight of each sample. Consequently, only four out of twelve samples contained approximately 5 ng of Zn, whereas the other chondrule fractions contained $\sim$2 ng. These less concentrated aliquots were analyzed after repeat measurements of 5 ppb BHVO-2 (n=9) and bulk CV chondrite (n=12) solutions, to redetermine the analytical uncertainty of these smaller samples. The $\updelta$$^{66}$Zn values of the reference standards are reproduced accurately, albeit with a relatively larger uncertainty for the CV chondrite ($\updelta$$^{66}$Zn$_{BHVO-2}$ = 0.27$\pm$0.05 $\permil$ [2SD], $\updelta$$^{66}$Zn$_{CV}$ = 0.27$\pm$0.19 $\permil$ [2SD]), whereas the $\updelta$$^{68}$Zn values are less accurate and precise ($\updelta$$^{68}$Zn$_{BHVO-2}$ = 0.34$\pm$0.13 $\permil$ [2SD], $\updelta$$^{68}$Zn$_{CV}$ = 0.49$\pm$1.17 $\permil$ [2SD]). The 2SD error of the CV chondrite for a 5 ppb solution is more realistic than the small error found for similarly sized BHVO-2 aliquots, since we measure with half the signal intensity. The good reproducibility of BHVO-2 is likely related to optimal circumstances for instrumental drift. Most of the individual measurements are not mass-dependent due to the lesser precision of the $\updelta$$^{68}$Zn values, most likely because of the lower signal/noise ratio of the analyses. Nevertheless, our $\updelta$$^{66}$Zn standard data are reproducible within an uncertainty of 190 ppm and in excellent agreement with literature values of BHVO-2 and bulk CV chondrites. For this reason, we report only the $\updelta$$^{66}$Zn values of the chondrule fractions, although $\updelta$$^{68}$Zn values can be found in Table \ref{tab:LeovilleZn}. The chondrule cores have $\updelta$$^{66}$Zn values ranging between --0.36$\pm$0.12 $\permil$ and --0.54$\pm$0.19 $\permil$, and are tightly clustered with an average of --0.43$\pm$0.14 $\permil$. These values coincide with the most negative whole chondrule data (e.g. core and rim) obtained for Allende \citep{Pringle2017}. The igneous rims have more positive $\updelta$$^{66}$Zn signatures, ranging between --0.14$\pm$0.19 $\permil$ and 0.23$\pm$0.19 $\permil$, with an average of --0.01$\pm$0.30$\permil$. We note that the sample with the outlying value of 0.23$\permil$ (Ch5) has a distinct rim mineralogy compared to the other chondrules, olivine being absent (Fig. \ref{fig:MapsRims}). The two matrix analyses from locations surrounding Ch1 and Ch2 yield similar Zn isotope signatures with $\updelta$$^{66}$Zn values of 0.24$\pm$0.12 $\permil$ and 0.14$\pm$0.12 $\permil$, respectively. These values coincide with the average bulk $\updelta$$^{66}$Zn value of CV chondrites (0.24$\pm$0.12 $\permil$, \citealp{Pringle2017}), but are somewhat lower than the matrix-rich aliquot taken from Allende (0.35 $\permil$, \citealp{Pringle2017}).

\section{Discussion}

\subsection{Size limitations on Zn isotope analyses}
\label{limitations}
\begin{figure}[]
	\centering
	\includegraphics[width=1.0\textwidth]{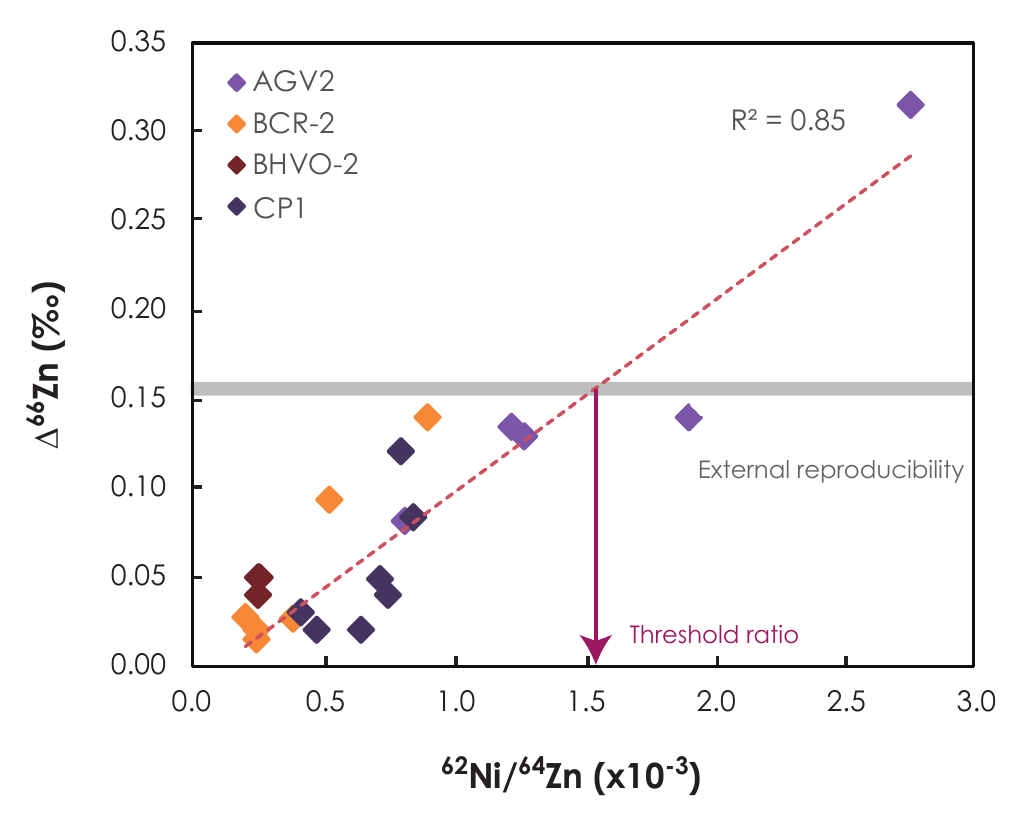}
	\caption{The deviation from the Zn mass-dependent fractionation line ($\Delta$$^{66}$Zn $\permil$) versus the $^{62}$Ni/$^{64}$Zn ratio of analyzed terrestrial standards by MC-ICPMS after correction of the $^{64}$Ni interference on $^{64}$Zn. The positive correlation through the data is interpreted as matrix effects from contaminants in the Zn-cut that affect the mass-dependency of the measurements. We show that we can apply a threshold Ni/Zn ratio to the data, above which the quality of the data is poorer than the external reproducibility. 
		\label{fig:NiZnthreshold}}
\end{figure}

\begin{figure}[]
	\centering
	\includegraphics[width=1.0\textwidth]{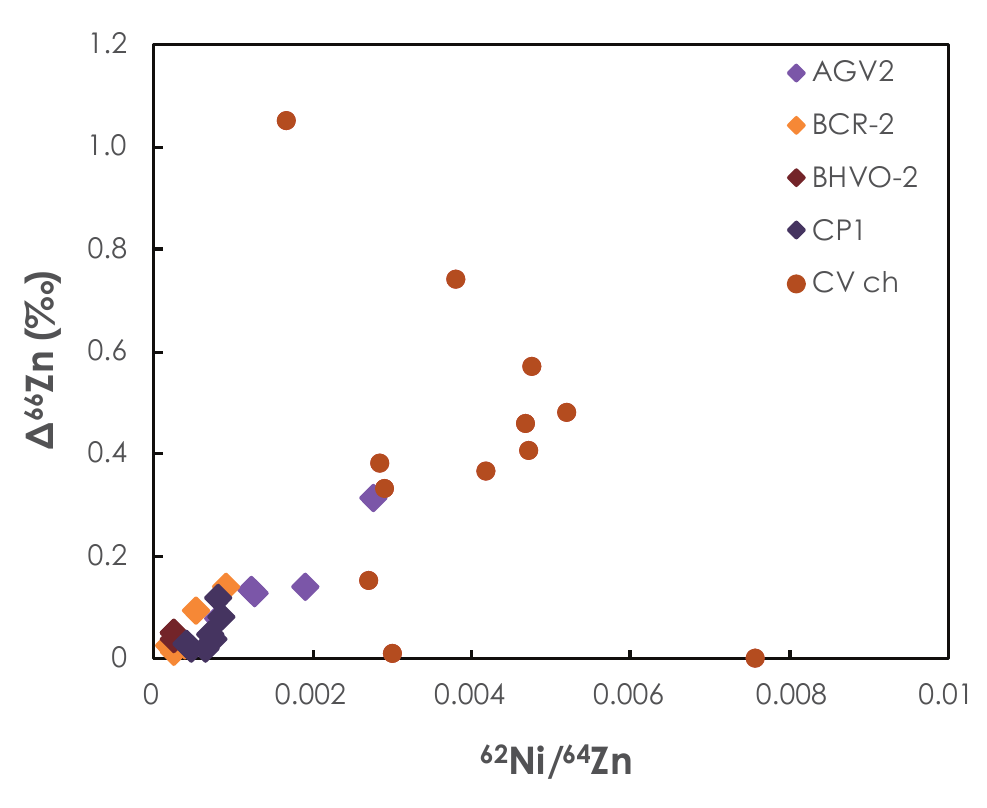}
	\caption{The same Figure as Figure \ref{fig:NiZnthreshold}, but including Leoville chondrule data from this study.
		\label{fig:NiZnthresholdchondrules}}
\end{figure}

Repeat analyses of 5 ng Zn aliquots from the BHVO-2 standard yield accurate and reproducible $\updelta$$^{66}$Zn values, in agreement with literature data. The external reproducibility of 80 ppm (2SD) on $\updelta$$^{66}$Zn is identical to the spread found in inter-laboratory analyses of BHVO-2. This error is lower than the reproducibility of separate digestions of geological standards, each containing 5 ng of Zn (120 ppm on $\updelta$$^{66}$Zn). The higher error can be explained in two ways. First, we find a positive correlation between the $^{62}$Ni/$^{64}$Zn ratio and the deviation from the mass-dependent fractionation line for terrestrial standards (Fig. \ref{fig:NiZnthreshold}). Since we correct for the interference of $^{64}$Ni on $^{64}$Zn by monitoring the abundance of $^{62}$Ni \citep{Moynier2015}, the $^{62}$Ni/$^{64}$Zn ratio likely reflects the abundance of residual matrix elements in the Zn-cut after purification, which will introduce matrix effects during Zn isotope analyses. This ratio appears to be independent from the initial Ni/Zn ratio before purification. For example, the Ni/Zn ratio of AGV2 is a factor 2 lower (0.22) than that of CP1 (0.42), but has higher residual Ni/Zn ratios. Moreover, AGV2 has an intermediate Zn concentration (86 ppm) relative to CP1 (39 ppm) and BCR-2 (127 ppm). Hence, it is unclear how the matrix composition of AGV2 would affect the deviation from the mass-dependent fractionation line. It is possible that during or after column purification the AGV2 samples were contaminated somehow, since it would only take $\sim$10 pg of added Ni to contaminate the sample. Nevertheless, future Zn isotope analyses on terrestrial samples with similarly small sample sizes would benefit from the dependency of the $^{62}$Ni/$^{64}$Zn ratio on a successful outcome of a measurement. A threshold ratio ($^{62}$Ni/$^{64}$Zn $<$ 1.5 $\times$ 10$^{-3}$) could be assigned to ensure measurements fall within the external reproducibility found in this study. This way, before all sample is consumed during analyses, another column pass might be performed to remove the residual matrix. However, this dependency is likely not relevant for Leoville fractions analyzed in this study. The lack of correlation between $^{62}$Ni/$^{64}$Zn and the mass-dependency in these samples (Fig. \ref{fig:NiZnthresholdchondrules}) suggests that the large variations between $\Delta$$^{66}$Zn (= $\updelta$$^{68}$Zn/2-$\updelta$$^{66}$Zn, the deviation from the mass-dependent correlation line) with similar $^{62}$Ni/$^{64}$Zn ratios are probably the result of low signal/noise ratios on $^{68}$Zn for samples $<$5 ng Zn (see below).  \\

Another caveat in analyzing very small Zn aliquots is matrix effects from resin-derived organics during mass spectrometry measurements. Anion exchange resins have been shown to contaminate samples with organics stripped from the resin, even after extensive cleaning \citep{Pietruszka2008}. These organics produce matrix effects during MC-ICPMS measurements. Although our purification protocol significantly reduces the amounts of acids used, the relative column size is not significantly adjusted to the smaller samples. The samples in this study are 2-3 orders of magnitude smaller than previous studies, whereas the reduction of resin volume is only by a factor 3.5. Hence, the resulting Zn-cut contains a relatively large proportion of organics that can affect our data. Zinc isotope analyses of BCR-2, AGV2 and CP1 show that a small offset may exist between our dataset and reported values from literature ($<$0.08 $\permil$) even though our data are within error of these reported values and the offset is within the external reproducibility of our measurements. This potential negative offset is likely caused by matrix effects from resin-derived organics. Both the Zn procedural blank and the organics can be limited by decreasing the total column volume of the Zn purification protocol. Here, we are restricted by the amount of sample that can be passed over the column without stagnating the resin. The processing of even smaller samples may result in a relative increase of the residual matrix/Zn ratio (Fig. \ref{fig:FigZncutcontamination}), requiring another column pass, which in turn would result in a larger organics/Zn ratio. Hence, a sample size of 5 ng Zn is probably at the limit of what is possible at the moment for reasonable Zn isotope data, reasonable meaning that the external reproducibility of our measurements is relatively small compared to the expected range of $\updelta$$^{66}$Zn values ($\sim$2 $\permil$, \citealp{Moynier2017}).\\

Our aim in this study was to proof the concept of measuring very small Zn aliquots by sampling fractions of Leoville chondrules. As reported in section \ref{Results_Znisotopes}, most chondrule fractions contained $<$5 ng of Zn. For most of the individual measurements of these fractions, we cannot test the mass-dependency due to the lesser precision of the $\updelta$$^{68}$Zn values, most likely because of the lower signal/noise ratio of the analyses. Although our standard data using such small Zn concentrations reflect accurate values for $\updelta$$^{66}$Zn, the Leoville chondrule fractions may suffer from a relatively large blank contribution, which is based on our repeated blank tests conservatively estimated to be $<$0.935 ng, but on average $<$0.350 ng. This could contribute up to 50\% of the total sample size. Hence, we may expect a maximal positive shift in $\updelta$$^{66}$Zn values of $<$0.36 $\permil$ for the chondrule cores, $<$0.15 $\permil$ for the igneous rims and $<$0.05 $\permil$ for the matrix. This correction uses $\updelta$$^{66}$Zn = 0 $\permil$ for the blank, which is based on measurements of the isotope composition of the procedural blank in this study. To this end, we have put four aliquots of 1 ng JMC Lyon standard through our Zn purification procedure and have measured these separately for their Zn isotope composition against a pure and unprocessed JMC Lyon standard (Table \ref{tab:standardsZn}). The $\updelta$$^{66}$Zn values of the processed standards are all indistinguishable from the unprocessed standard, which suggests that the composition of the blank is identical. This is in agreement with similar Zn isotope measurements of bulk column blanks from Zn purification protocols using the same anion resin and reagents \citep{Shiel2009}. In detail, these authors have processed bulk column blanks of Zn elutes from anion resins (AG MP1, 200-400 mesh, macroporous version of AG1) and added that to a pure Zn standard solution (in-house standard PCIGR1, identical in Zn isotope composition to the JMC-Lyon standard used in this study; \citealp{Shiel2009}) at different ratios between blank and standard (e.g., 0.2, 10 and 50 \%). \citet{Shiel2009} show, using the same HNO$_{3}$ flux sample treatment after purification as in this study, that the blank contaminated Zn standards are typically positive and within 2SD error ($\updelta$$^{66}$Zn = 0.1 $\permil$) of the pure Zn standard. Hence, the negative $\updelta$$^{66}$Zn values observed for the chondrule cores are not likely to be the result of a blank contribution with a very anomalous Zn isotope values. It is probable that the chondrule cores were somewhat more negative than we observe here. Furthermore, we note that the chondrule cores, although measured for different sample sizes (e.g., 1-3 ng Zn), exhibit identical $\updelta$$^{66}$Zn values (--0.43$\pm$0.14 $\permil$, 2SD), suggesting that the upper limit blank contribution of 0.9 ng Zn is an overestimation and that it is more realistic to use the average blank of 0.3 ng Zn or lower. Thus, such a blank contribution would yield corrected $\updelta$$^{66}$Zn values of --0.56 $\permil$ for the chondrule cores, which is within error of our measurements. Hence, we suggest that the Zn isotope data from the chondrule cores is accurate enough to allow for cosmochemical interpretations, even though being less robust than 5 ng Zn samples due to a relatively higher blank contribution and a lower signal/noise ratio during MC-ICPMS measurements. \\
The applied blank correction may be counteracted (for negative $\updelta$$^{66}$Zn values) or enlarged (for positive $\updelta$$^{66}$Zn values) due to organic interferences. However, since the Zn isotope data for different chondrule fractions (cores, igneous rims and matrices) are tightly clustered within each group (using different Zn concentrations during isotope analyses) and the procedural blanks are variable, we suggest that this offset is likely minor. \\

In summary, we show that Zn isotope analyses are size limited by matrix elements contaminating the Zn-cut, by resin derived organics, low signal/noise ratios and blank contributions. Samples $>$5 ng Zn result in mass-dependent, accurate and reproducible data when purified sufficiently, whereas samples $<$5 ng are mostly susceptible to low signal/noise ratios of $^{68}$Zn and relatively high blank contributions.
  
\subsection{Petrological and compositional relationships between Leoville fractions}
\label{elements_complementarity}
To understand the magnitude and direction of Zn isotope fractionation between cores, igneous rims and surrounding matrix within the Leoville CV3.1 chondrite, it is first important to discuss the petrological and compositional context of the chondritic fractions and how they complement each other. 

\subsubsection{Compositional heteorogeneity of CV chondrite matrix}
\begin{sidewaysfigure}[]
	\centering
	\includegraphics[width=1.1\textwidth]{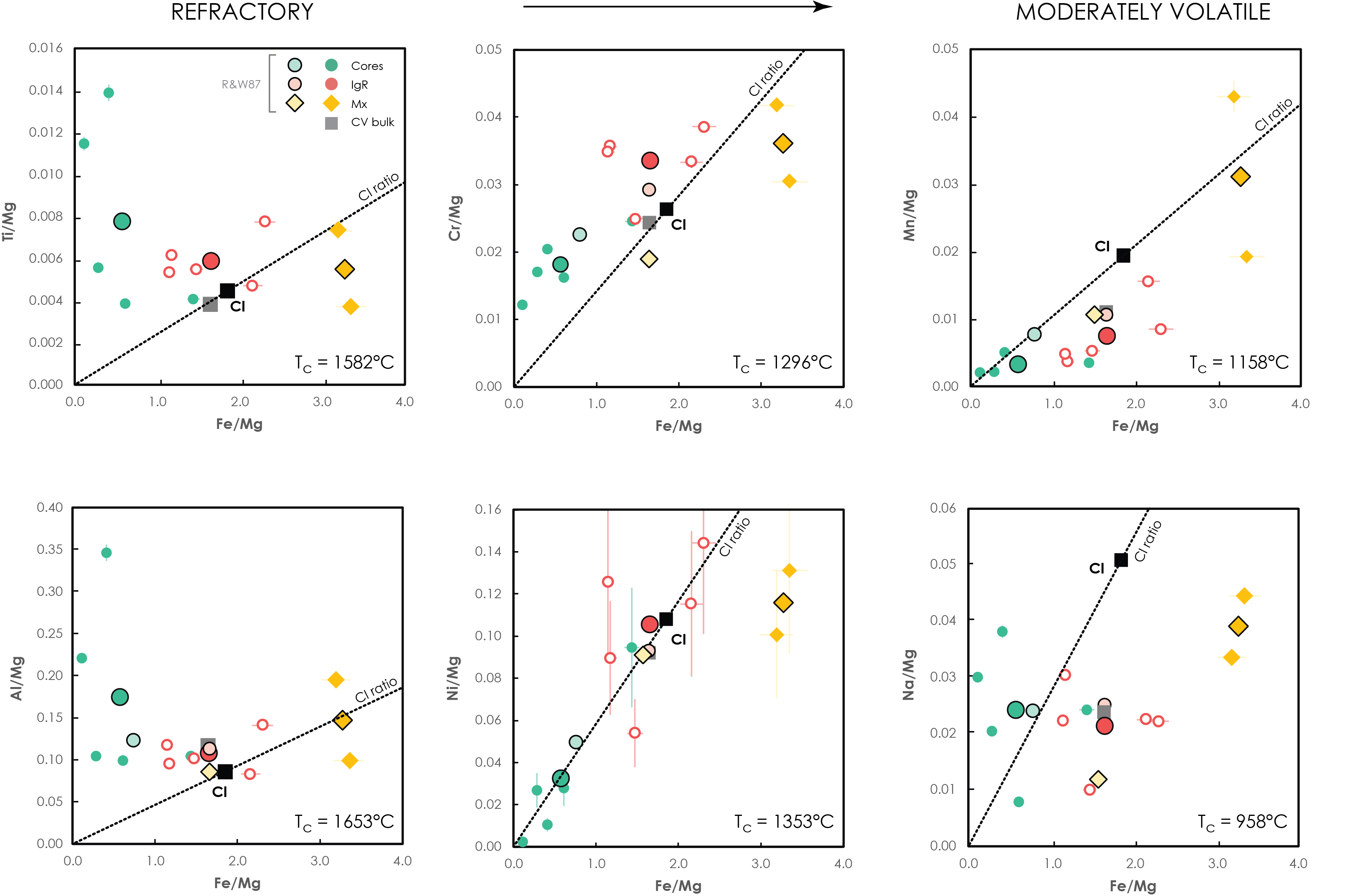}
	\caption{Plots of elemental ratios normalized against Mg with increasing 50\% condensation temperatures from left to right \citep{Lodders2003} for Leoville chondrule cores, igneous rims and matrices. The bulk compositions of bulk CI and CV chondrites are also plotted, along with the CI ratio correlation often shown in complementarity related plots.  
		\label{fig:Majorelements}}
\end{sidewaysfigure}

\begin{figure}[]
	\centering
	\includegraphics[width=1.0\textwidth]{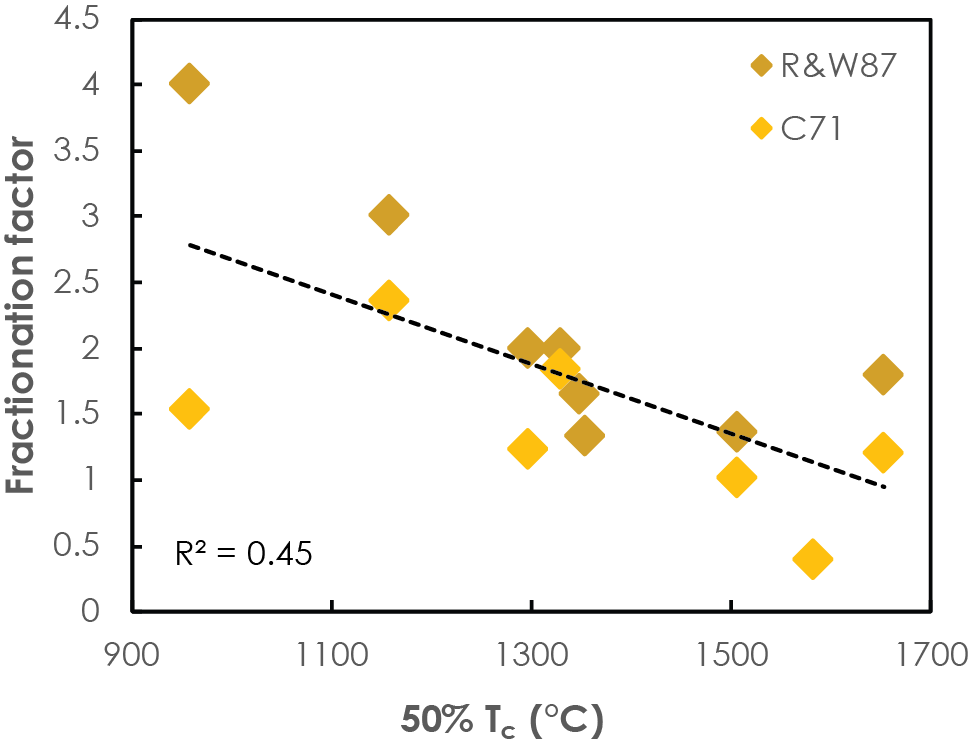}
	\caption{Matrix fractionation factors calculated from elemental ratios normalized to Mg from this work, \citet{Rubin1987} (neutron activation analyses) and \citet{Clarke1971} (wet chemical analyses) against 50\% condensation temperatures \citep{Lodders2003}. The dashed regression line is taken from all data. RW87 = \citet{Rubin1987} and C71 = \citet{Clarke1971}.
	\label{fig:FractionationCondensation}}
\end{figure}

\citet{Rubin1987} were the first to measure the composition of multiple chondrule cores and their igneous rims within a CV chondrite (e.g., Allende). We show in Figure \ref{fig:Majorelements} that the average core and rim compositions in this study closely match the average data of \citet{Rubin1987}. In the same study, bulk matrix compositions of Allende were measured using various analytical techniques, including defocused beam, neutron activation and wet chemical analyses. The Allende matrix composition differs significantly from our Leoville matrix analyses in that the latter is relatively enriched in volatiles. We observe a tentative correlation between the 50\% condensation temperature and the fractionation factor between our dataset and that of \citet{Rubin1987} and \citet{Clarke1971} using neutron activation and wet chemical analyses, respectively (Fig. \ref{fig:FractionationCondensation}). Note that we do not compare our data to defocused beam analyses, since these type of measurements may cause artificial variability in Mg amongst others \citep{Barkman2013,Zanda2018}. The Fe/Mg ratio in the Leoville matrix is a factor $\sim$2 higher (T$_{c}$$<$1328$^{\circ}$C), whereas the Mn/Mg ratio is a factor $\sim$3 higher (T$_{c}$$<$1158$^{\circ}$C) and the Na/Mg ratio a factor $\sim$4 higher (T$_{c}$$<$958$^{\circ}$C) compared to the Allende matrix (Fig. \ref{fig:Majorelements}). Defocused beam analyses of CV chondrite matrix including Leoville show that average matrix compositions are typically super-chondritic and fairly uniform within the same chondrite (Table \ref{tab:Matrix}, \citealp{Clarke1971,McSween1977,Kracher1985,Zolensky1993,Klerner2001,Huss2005,Hezel2010,Palme2015}). Even considering a $\sim$40\% offset between microprobe and bulk analyses of Orgueil, the latter is more Fe/Mg rich \citep{Zanda2018}, suggesting that Fe/Mg ratios measured by microprobe are underestimated.  On a sub-mm scale, the variations in Fe/Mg become larger and matrix compositions as measured in this study are not uncommon \citep{Palme2015}. Individual point analyses of matrix in Efremovka and Mokoia have Fe/Mg and Mn/Mg ratios $<$4.4 and $<$0.045, respectively. Since our matrix samples have been sampled on a sub-mm scale, it is not suprising to find such highly super-chondritic values. The small scale matrix variations could be the result of nebular heterogeneity \citep{Grossman1996} or redistribution through parent body alteration, since it has been suggested that the most unaltered chondrites have CI-like matrix compositions \citep{Zanda2018}. During aqueous alteration, formation of secondary calcite, fayalitic olivine and phyllosilicate/sulfide compounds will redistribute elements throughout the matrix \citep{Wasson2009}. For example, the analyses of a matrix region enriched in secondary calcite, would be defined by higher Ca/Mg ratios, whereas regions where sulfide precipitated are characterized by higher Fe/Mg and S/Mg ratios. Contrary, our matrix analyses show a fairly continuous trend between elemental condensation temperature and fractionation (Fig. \ref{fig:FractionationCondensation}), which cannot easily be explained by redistribution through aqueous alteration. Moreover, since the matrix regions sampled from our Leoville section show very limited evidence for alteration (e.g., no observations of larger secondary carbonate or sulfide agglomerates), we suggest that our matrix data mainly reflect nebular heterogeneity, rather than variability through aqueous alteration. This is likely true at least for refractory and moderately volatile elements. More volatile elements such as Na, K and S have been shown to transport into chondrules during very early stages of thermal metamorphism, during which the matrix is being depleted and the chondrules enriched \citep{Grossman2005}. This is in agreement with the Leoville matrix being lower in Na/Mg relative to the CI ratio, whereas the chondrule cores have higher values (Fig. \ref{fig:Majorelements}). For other moderately volatile elements, the matrix composition is close to the CI ratio. Moreover, the expected fractionation factor for the Na/Mg ratio between Allende and Leoville matrix is offset from the general trend (Fig. \ref{fig:FractionationCondensation}). For moderately volatile elements with 50\% condensation temperatures $>$1000 $^{\circ}$C, we suggest that the Leoville matrix analyses reflect primary nebular compositions. Thus, we can use our data to infer genetic relationships between chondrule cores, igneous rims and the surrounding matrix.

\subsubsection{Complexity of chondrule petrology}
Chondrules are typically divided into type I (FeO-poor) and type II (FeO-rich) compositions, after which they are further classified based on porphyritic, barred or cryptocrystalline textures and mineralogies (olivine and/or low-Ca pyroxene phenocrysts). Most chondrules have experienced more than one (strong) melting event, as evidenced by primary and secondary cores (separated by a metal/sulfide rim, generally with the same mineralogy and oxidation degree) and surrounding igneous rims \citep{Rubin2010,Scott2014}. These rims are usually finer-grained ($<$100 $\upmu$m) than their corresponding cores, even though they are often described as 'coarse-grained' to avoid confusion with fine-grained matrix rims. Igneous rims have so far been observed enveloping chondrule cores in CV ($\sim$50 \%, \citealp{Rubin1984,Rubin1987}), CO ($<$1 \%, \citealp{Rubin1984}), CR (unknown, \citealp{Krot2004}) and ordinary chondrites ($\sim$10 \%, \citealp{Rubin1984, Krot1995}). \citet{Rubin1987} described the petrology and composition of CV chondrite cores and corresponding igneous rims and found that, similar to our chondrule data, individual core-rim pairs were not compositionally matched and probably did not form during the same heating event. It is generally acknowledged that igneous rims formed after the melting and solidification of the host chondrule \citep{Krot1995}, unlike observed mineralogical zonation of chondrule cores that have displayed open system behaviour (i.e., low-Ca pyroxene rims, \citealp{Friend2016,Barosch2019}). The rims had to have been heated to lower temperatures (T$_{sol}$$<$1000-1200 $^{\circ}$C, \citealp{Hewins2005,Jones2018}) than the cores (T$_{liq}$$<$1700$^{\circ}$C), to avoid complete melting and homogenization of the chondrule \citep{Rubin1987}. Thus, the igneous rims may have experienced (liquid-solid) sintering rather than complete melting, considering the liquidus temperature. This begs the question whether igneous rims formed from similar precursor materials as their host chondrules and if this precursor dust is similar in composition to the surrounding matrix. With this in mind, we will discuss below scenarios of complementarity and non-complementarity (for a detailed overview of both models see \citealp{Hezel2018,Zanda2018}) during the formation of the Leoville chondrule cores and rims. 

\subsubsection{Genetic relationships of Leoville fractions}
\begin{sidewaysfigure}[]
	\centering
	\includegraphics[width=1.0\textwidth]{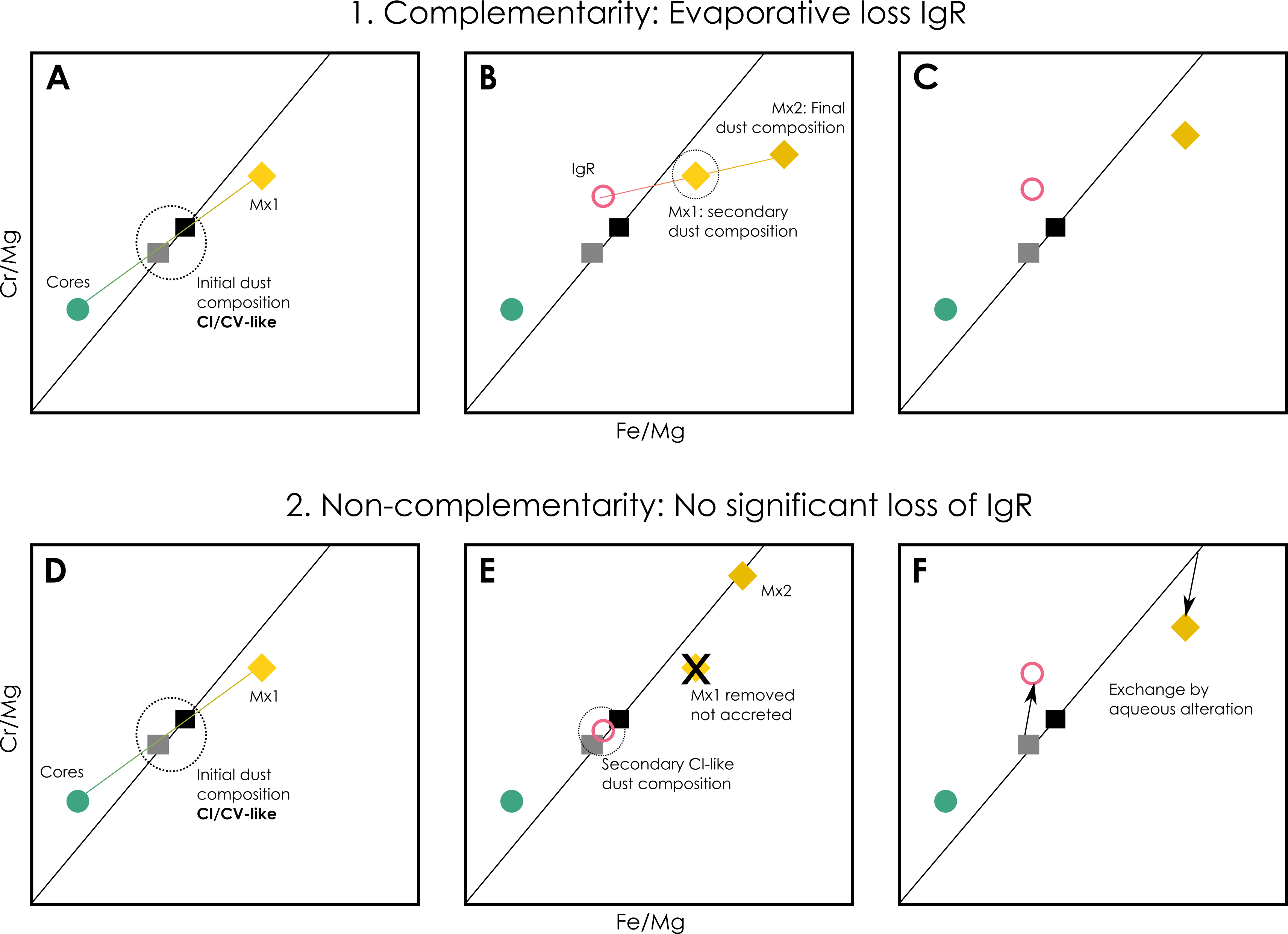}
	\caption{Different major element scenarios for complementarity and non-complementarity between Leoville cores and igneous rims, as well as igneous rims and surrounding matrix. Mx1 is the primary residual matrix after formation of the chondrule cores, whereas Mx2 is the residual matrix after formation of the igneous rims. In the complementarity scenario, Mx1 is the precursor material to form the igneous rims (having experienced evaporative loss) and the complementary volatile-rich Mx2. In the non-complementarity scenario, the precursor material to both chondrule cores and igneous rims is chondritic and not genetically related to the chondrules. In this model, the igneous rims experience limited evaporative loss and maintain their intial chondritic composition.
		\label{fig:Scenarios}}
\end{sidewaysfigure}

\begin{sidewaysfigure}[]
	\centering
	\includegraphics[width=1.1\textwidth]{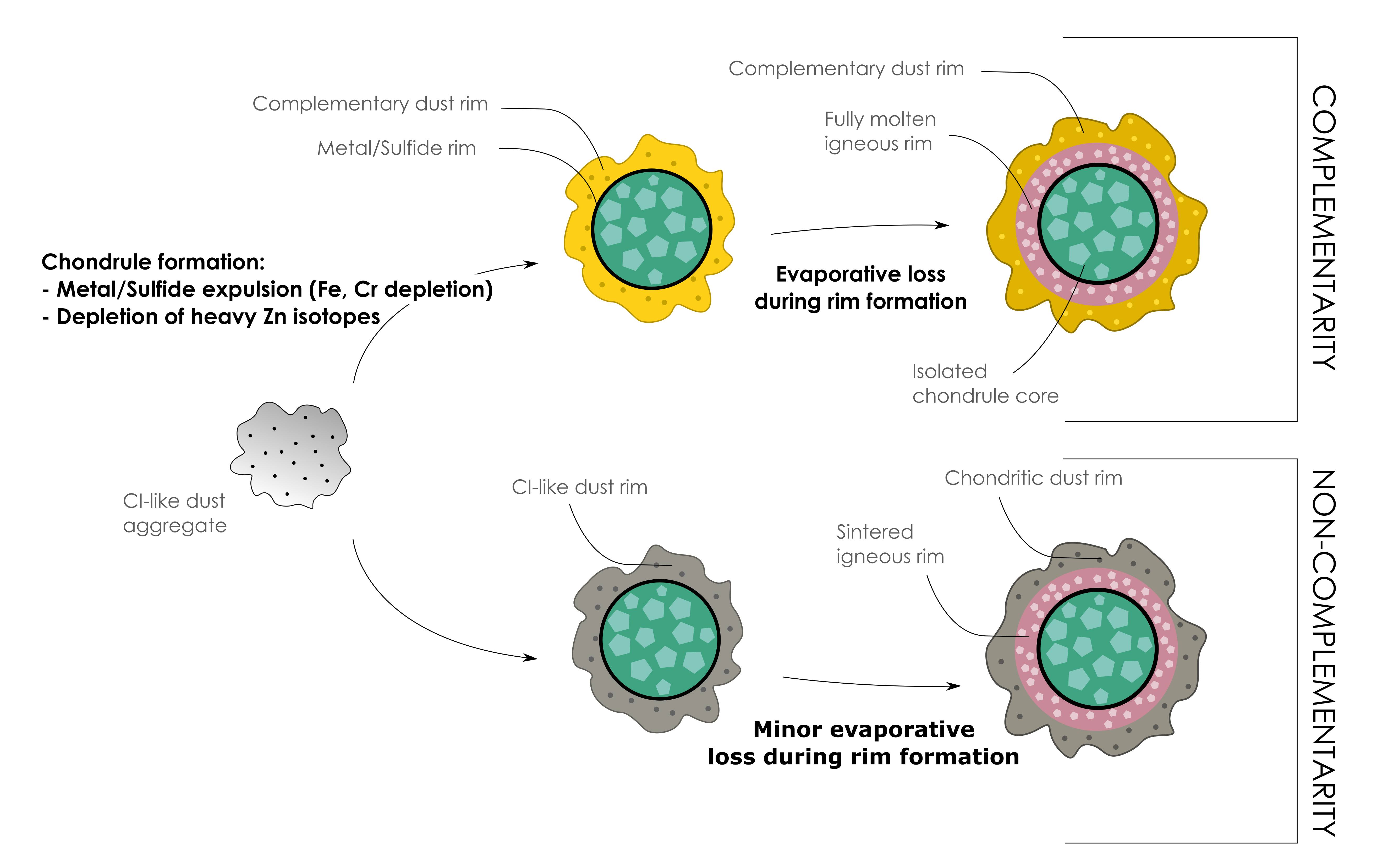}
	\caption{A schematic overview of chondrule formation in the complementarity and non-complementary scenarios from Figure \ref{fig:Scenarios}, using the same color coding. 
		\label{fig:SchematicChFor}}
\end{sidewaysfigure}

In recent complementarity studies that assess the genetic relationship between chondrules and matrix \citep{Bland2005,Palme2015,Hezel2018}, chondrules are considered as single entities that have experienced one melting event, whereas in reality most chondrules likely experienced multiple heating events \citep{Bollard2017}. As such, the average composition of a chondrule core and its igneous rim cannot be complementary to the surrounding matrix. It is, however, possible to evaluate potential complementarity between the core and rim as well as the rim and the surrounding matrix. Here, we discuss the scenarios of complementarity and non-complementarity for our petrological observations as well as major and minor element data. In Figures \ref{fig:Scenarios} and \ref{fig:SchematicChFor} we present these scenarios with Cr/Mg versus Fe/Mg ratios as an example. Complementarity between chondrule and matrix requires the precursor material to have a CI-like composition (e.g., falling along the CI ratio line, Fig. \ref{fig:Scenarios}) and the chondrules and matrix to have formed in the same reservoir \citep{Hezel2018}. If we take the average chondrule core composition as a starting point, the resulting primary dust rim around the core (from which the igneous rim formed) must be equal to Mx1 if the ratio between chondrule and dust rim was approximately 1:1. A calculation (using Adobe Illustrator PathArea plugin) of the surface areas of chondrule cores and igneous rims from the five chondrules analyzed in this study shows that this ratio is realistic. If the primary dust rim subsequently experienced another heating event and evaporative loss, then the final composition of the igneous rim could be near chondritic (again with a dust:chondrule ratio of 1:1), whereas the final matrix (Mx2) would be super-chondritic. While our data appear to be perfectly in agreement with the complementarity scenario, applying the same model to other major element plots from Figure \ref{fig:Majorelements} yields very different chondrule:dust ratios. For example, using refractory elements Ti and Al would result in a near 100 \% abundance of Mx1, whereas the Na/Mg ratios requires a near 100\% abundance of chondrules. In the case of Ni/Mg and Mn/Mg it is not even possible to produce a Mx1 that lies between the igneous rim and Mx2 composition, when the precursor dust has to have a CI-like composition. Similar discrepancies in chondrule/matrix ratios have been previously observed for CR and CO chondrite assemblages \citep{Zanda2018}. \\

In a non-complementarity scenario, such as envisioned by \citet{Zanda2018}, the final matrix composition of a chondrite is CI-like and chondrules are not produced in the same chemical reservoir. Any deviations from a CI-like matrix composition are considered to be the result of aqueous alteration processes. If we consider a chondritic starting composition for the primary dust rim accreted on the chondrule cores (Figs. \ref{fig:Scenarios} and \ref{fig:SchematicChFor}), then the heating event that produced the igneous rims required thermobaric conditions that evaded significant loss of non-refractory elements such as Fe, Cr, Ni and Mn, since the final composition of the igneous rims is also chondritic. The matrix complementary to the chondrule cores (Mx1) must have been removed from the CV chondrule reservoir, or the chondrules were transported to a different disk region. While this scenario works demonstrated solely by the elemental ratios of cores, rims and matrix, it is difficult to explain the petrological features of the igneous rims (i.e., relict forsterite).\\

Alternatively, in a scenario adopted from \citet{Marrocchi2013} for the formation of pyroxene-porphyritic (PP) chondrules in Vigarano, the matrix material that was present during chondrule core formation was thermally processed alongside the cores. Depleted forsterite-rich dust was subsequently accreted to the chondrule cores, which could have interacted with a complementary volatile-rich gas, bringing the composition of the chondrule rims to (near-)chondritic \citep{Marrocchi2013}. In this scenario, the S, FeO and SiO-rich gas would have been reabsorbed into the silicate melt of the rims and would have co-crystallized sulfides and low-Ca pyroxene at sulfur saturation level. The resulting petrological features of this model are in agreement with our observations, namely the presence of relict forsterite grains overgrown by low-Ca pyroxene co-existing with troilite/metal assemblages. Moreover, this model also fits with the near-chondritic compositions of the igneous rims. We note that this model implies non-complementarity between chondrule cores and igneous rims in the classical sense \citep{Hezel2018}, but necessitates formation of both components in the same chemical reservoir. In section \ref{chondrule_formation}, we discuss these scenarios using Zn isotope analyses.

\subsection{The significance of Zn isotope fractionation in Leoville}
\label{significance}
The Zn isotope compositions of the Leoville chondrule cores and rims correspond to the most negative and most positive $\updelta$$^{66}$Zn values measured for Allende chondrules, respectively (Fig. \ref{fig:ChondruleZnisotopes}, \citealp{Pringle2017}). Although the petrology of the analyzed Allende chondrules is unknown, they likely reflect total Zn isotope signatures of combined cores and igneous rims. Hence, we suggest that the variable signatures of these chondrules ($\updelta$$^{66}$Zn = --0.45 - 0.18 $\permil$) reflect a mixing line between negative chondrule cores and more positive igneous rims of CV chondrules. Consequently, the $\updelta$$^{66}$Zn range observed for Allende (an oxidized $>$CV3.6 chondrite) chondrules is equivalent to the range observed for Leoville (reduced CV3.1) chondrules, suggesting that the degree of thermal metamorphism experienced by Allende is inconsequential for the Zn isotope variability in its chondrules. This suggests that the exchange of Zn between chondrules and matrix during thermal alteration must have been limited. However, Zn is considered to be highly mobile in aqueous fluids \citep{Pringle2017,Zanda2018} and the effect of aqueous alteration on the Zn isotope composition of the Leoville chondrules must be discussed first. \\

\begin{figure}[]
	\centering
	\includegraphics[width=1.0\textwidth]{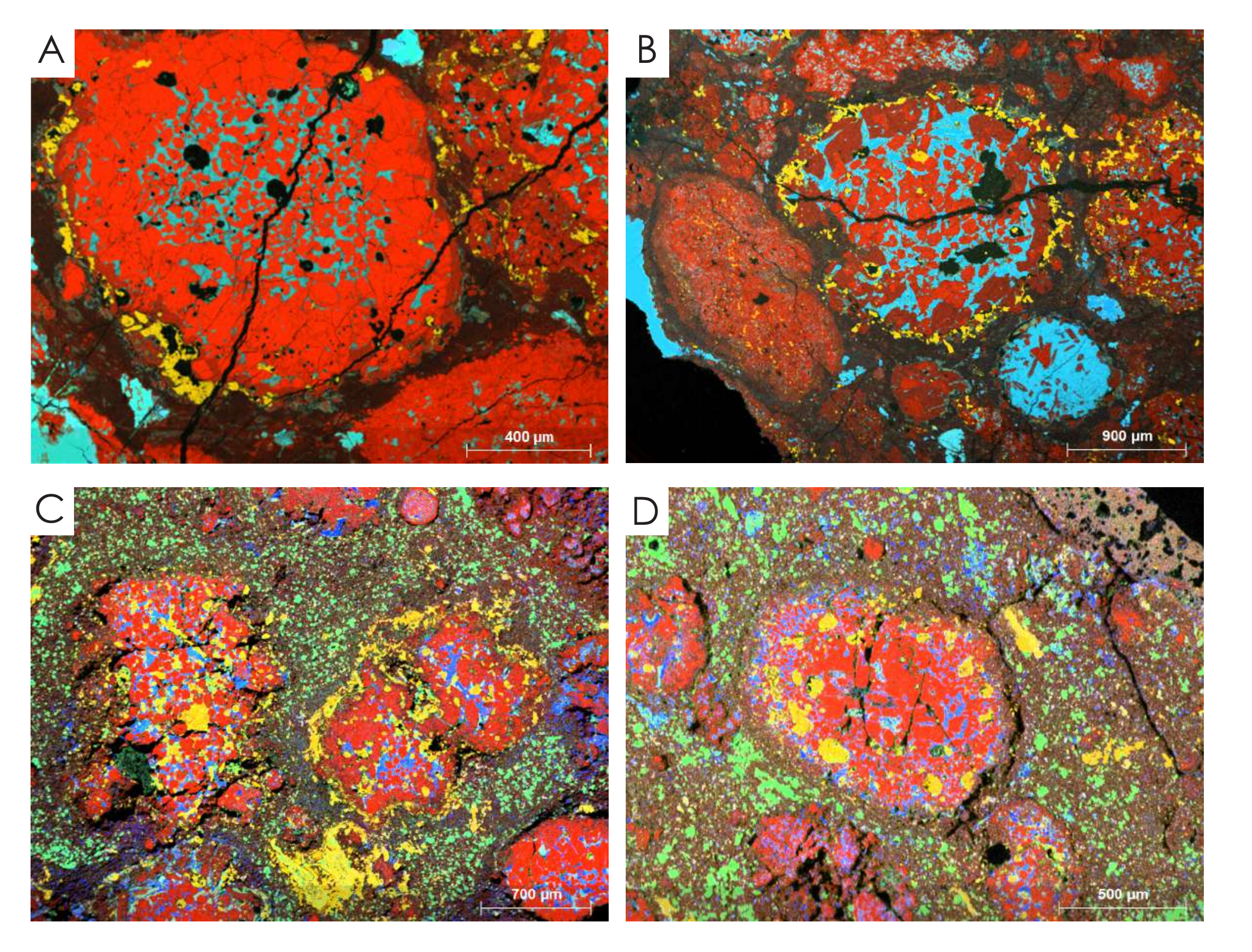}
	\caption{Mg (red), Ca (green), Al (blue) and S (yellow) elemental  maps of Leoville (A-B) and Allende (C-D) sections (chondrules in these images lie close to the ones sampled in this study). Note how the matrix of Leoville is mostly opaque and fine-grained, whereas the Allende matrix is coarse-grained with large sulfides and carbonates. The sulfides in Leoville are mainly present in the igneous rims or as sulfide rims to the chondrules, whereas in Allende, the sulfides have mobilized towards the interior cores in the most altered chondrules. 
		\label{fig:Allende}}
\end{figure}

Zinc isotope variations in Leoville and Allende chondrules are potentially the result of aqueous alteration. \citet{Alexander2019} has shown that the Zn concentration is correlated to the matrix abundance in carbonaceous chondrites, suggesting that chondrules were initially devoid of Zn. Thus, the limited alteration experienced by Leoville could have been sufficient to redistribute Zn from matrix to chondrules. It has been previously shown that other mobile elements such as Na and S can been extensively redistributed between chondrules and matrix in the most primitive ordinary chondrites  \citep{Grossman2005}. At the earliest stages of fluid-assisted metamorphism (petrological type 3.00-3.1), Na diffuses into chondrule mesostasis during vitrification and albite formation. At petrological types $>$3.1, Na transfers back to the matrix where feldspar starts to crystallize. Hence, the direction of Na mobilization is dependent on the sink of Na at different stages of thermal metamorphism. This raises the question what the sink of Zn is during these metamorphic stages, since Zn can act as a lithophile as well as a chalcophile element. Zinc is chalcophile and stored in the sphalerite component of sulfides in the least altered chondrites Semarkona (LL3.00), ALH 77307 (CO3.0) and Kainsaz (CO3.1) \citep{Johnson1991}. Any petrological grade higher than Semarkona in ordinary chondrites has detectable and increasing amounts of Zn incorporated in chromites that have crystallized or re-equilibrated during thermal metamorphism \citep{Johnson1991,Chikami1999}. Hence, mild reheating in chondrites could redistribute Zn, similarly to Na and S. Spinel analyses from Allende chondrules reflect considerable ZnO concentrations ($<$0.55 wt\%; average = 0.28 wt\%; \citealp{Riebe}). The abundance of spinel in Allende chondrules is estimated to be 0.6 wt\%, resulting in an average Zn contribution of $\sim$13 ppm in Allende chondrules from spinels. Total Zn concentrations of chondrules analyzed by \citet{Pringle2017} range between 31 and 94 ppm, suggesting that the contribution from potentially secondary Zn in spinel does not account for the total Zn in chondrules. Hence, some Zn could be primary. However, without a detailed knowledge of the petrology of the Allende chondrules analyzed for Zn isotopes, we cannot say with any certainty to which extent the Zn in the chondrules is primary. Another potential source for secondary Zn could be derived from Fe sulfides redistributed to Allende chondrule interiors (Fig. \ref{fig:Allende}, \citealp{Zanda1995,Hewins1997,Grossman2005}). Unlike Leoville, the relatively high degree of thermal metamorphism in Allende has resulted in FeS coarsening and redistribution towards the chondrules cores in some of the more altered chondrules (Fig. \ref{fig:Allende}), suggesting that Zn redistribution in Allende chondrules has been more pervasive. Thus, in theory, Allende chondrules may have only contained secondary Zn. However, it is possible that the redistribution of sulfides and Zn in Allende chondrules would have taken place only within the chondrules, which would have acted as a closed system. In this case, FeS from the igneous rims would have been transferred to the chondrule interiors, but Zn would not have easily transferred from the matrix to the chondrules. This is a realistic possibility, since secondary minerals formed within the matrix (e.g., carbonates, sulfides, sulfates and phosphates) could have represented a considerable sink for Zn by itself, preventing redistribution towards the chondrules. This model would explain the similar range of $\updelta$$^{66}$Zn values observed in Leoville and Allende chondrules, in spite of their different alteration degrees. This model assumes that the sulfides in the igneous rims are primary unaltered components, not mobilized from the matrix. Indeed, similar to the CV3.1-3.4 chondrite Vigarano, the absence of magnetite-carbide-sulfide associations \citep{Krot2004} and the nature of metal-sulfide structures (i.e., the absence of sulfide rims around metal, \citealp{Marrocchi2013}), suggest that the troilite grains in Leoville igneous rims are primary high-temperature components \citep{Marrocchi2013}. Furthermore, even though very pristine ordinary chondrites show redistribution of Zn to spinels, the carbonaceous chondrite Kainsaz reports an absence of secondary Zn in its chondrules. This implies that the Zn mobility was different for carbonaceous chondrites and that Leoville, being of similar petrological degree to Kainsaz, may have avoided significant redistribution of Zn. This is certainly in agreement with the tightly clustered $\updelta$$^{66}$Zn values of the Leoville chondrule cores. We would predict more variable $\updelta$$^{66}$Zn signatures depending on the chondrule core sizes and mineralogy if these values were dependent on the redistribution of Zn. However, sampled chondrule cores vary in size by a factor $\sim$10 and the texture and mineralogy of the chondrules varies considerably as well. Moreover, sphalerite dissolution of Zn would yield unfractionated or heavy Zn signatures of the hydrothermal fluid (depending on pH and pCO$_{2}$ conditions, \citealp{Fujii2011}), whereas subsequent incorporation into spinels would result in even heavier Zn isotope values relative to sphalerite. In contrast, we observe negative $\updelta$$^{66}$Zn values for Leoville chondrule cores. In summary, we suggest that the Zn isotope data from the Leoville chondrules reflect primary rather than secondary signatures.

\subsection{Zn isotope behaviour during chondrule formation}
\label{chondrule_formation}
\begin{figure}[]
	\centering
	\includegraphics[width=1.0\textwidth]{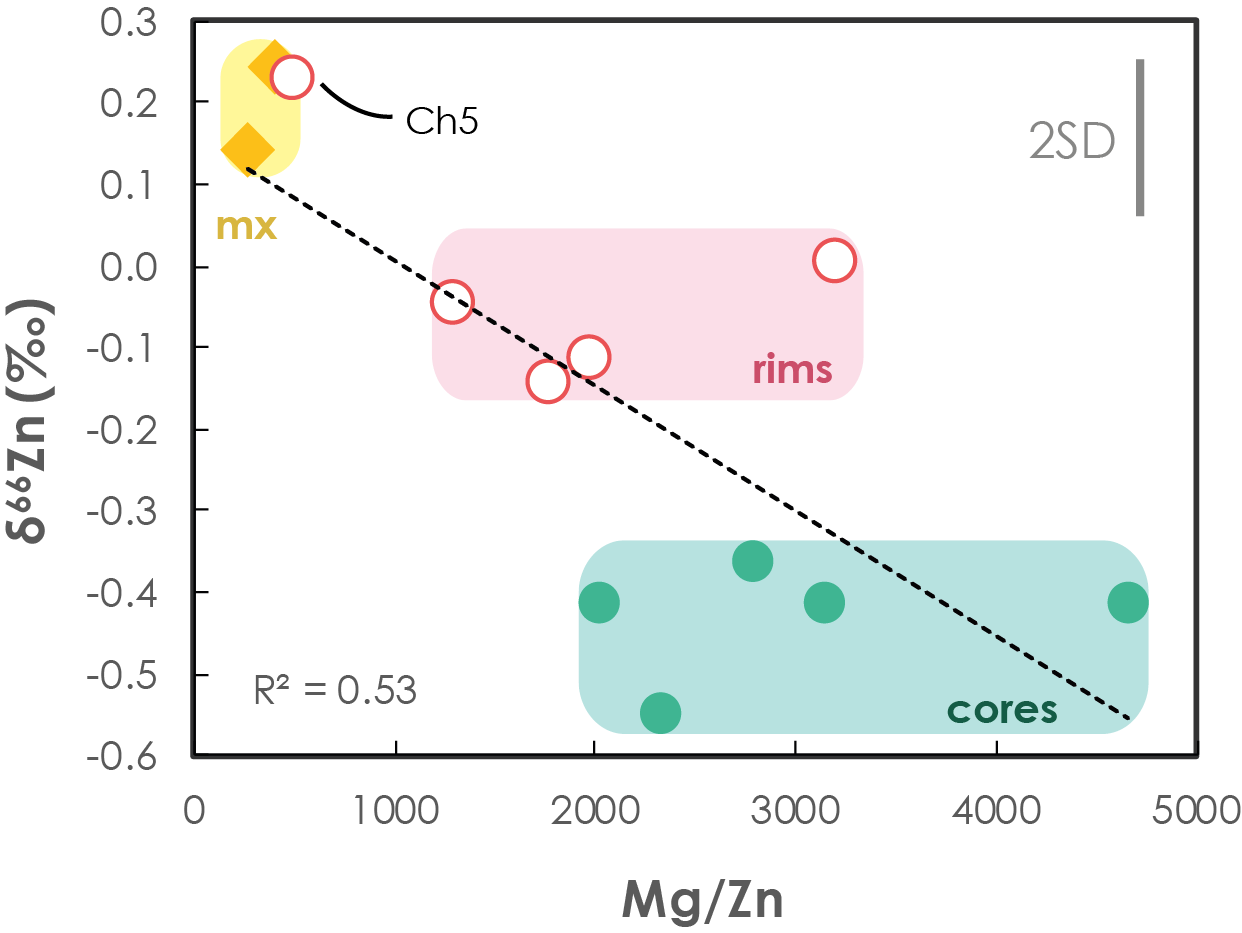}
	\caption{Zn isotope data of Leoville chondrule cores (closed green circles), igneous rims (open red circles) and matrix (yellow diamonds) against their Mg/Zn ratios. Mg concentrations have been obtained from ICPMS data and Zn concentrations are from MC-ICPMS analyses. Hence, the Mg/Zn ratios are relative and not absolute. The composition of Ch5 lies close to the matrix data and has a different mineralogy compared to the other igneous rims (see text for further explanation).
		\label{fig:ZnVsZnisotope}}
\end{figure}

The negative $\updelta$$^{66}$Zn values for chondrules relative to the CV bulk have been attributed to quantitative removal of sulfides with heavy Zn isotope signatures during chondrule formation \citep{Pringle2017}. This process results in a correlation between the Zn concentration and $\updelta$$^{66}$Zn values of the chondrules \citep{Pringle2017}. We observe a similar correlation between Mg/Zn ratios and $\updelta$$^{66}$Zn values of Leoville chondrule cores, igneous rims and matrix (Fig. \ref{fig:ZnVsZnisotope}). We note that the resulting regression may be scattered due to variable Mg concentrations between chondrule core and igneous rims \citep{Rubin1987}, which are generally higher for the cores than the rims. Nevertheless, the igneous rims have higher Zn concentrations and $\updelta$$^{66}$Zn values than their corresponding cores. Our data are in agreement with the heavy Zn isotope fraction being hosted by the Fe sulfides, since the igneous rims have a higher abundance of these components. As discussed in section \ref{significance}, these sulfides are likely primary features of the Leoville chondrules, developed during chondrule formation. Sulfide abundances in Vigarano CV chondrules have been linked to the co-crystallization of low-Ca pyroxene \citep{Marrocchi2013}. Hence, the higher the abundance of low-Ca pyroxene, the higher the sulfide concentration and predictably the higher the $\updelta$$^{66}$Zn value of the chondrule. Indeed, the igneous rim in Ch5, with a significantly higher abundance of low-Ca pyroxene relative to forsterite also accomodates the highest $\updelta$$^{66}$Zn value (Table \ref{tab:LeovilleZn}). The Ch1 igneous rim with the highest abundance of forsterite/enstatite has the lowest $\updelta$$^{66}$Zn value. Hence, our data is fully in agreement with the separation of silicate and sulfide melt during chondrule formation being at the core of Zn isotope fractionation. In this model, the chondrule cores must have lost their sulfide component to a considerable degree. This implies that the igneous rims maintained/retained a significant fraction of their sulfide inventory for them to show only slight depletions in $^{66}$Zn relative to the matrix. 

\citet{Marrocchi2013} have suggested that the formation of low-Ca pyroxene (PP) chondrules in Vigarano must have been the result of a reaction between partially depleted olivine-bearing precursors with a volatile (sulfur)-rich gas. Consequently, the PP chondrule composition becomes (near-)chondritic. This agrees with the observation of relict forsterite grains overgrown by low-Ca pyroxene as well as with the near-chondritic compositions of the Leoville igneous rims. These olivine-bearing precursors may represent initially CI-like dust that was thermally processed during the chondrule formation episode that formed the Leoville cores. If the volatile-rich gas reflects the complementary component of the simultaneously formed depleted olivine-bearing dust and the depleted type-I chondrule cores, then the cores and igneous rims are in fact genetically related and formed from the same chemical reservoir. In this scenario, the depleted olivine-bearing dust accreted onto the chondrule cores and subsequently reacted with a volatile-rich gas. The heavy Zn isotope fraction that was initially lost to the chondrule cores and the forsterite-rich thermally processed dust by sulfide vaporization, would have reabsorbed back onto the igneous rims together with S, FeO and SiO from the volatile-rich gas. The amount of interaction with this gas would have affected the Zn isotope composition of the rims (e.g., a higher reabsorbtion of Zn would yield a higher $\updelta$$^{66}$Zn value). The completion of the reaction from forsterite to low-Ca pyroxene yields $\updelta$$^{66}$Zn values similar to the matrix, suggesting that the initial composition of the dust that formed the chondrule cores and igneous rims was similar to the thermally unprocessed surrounding matrix. Hence, the Zn isotope data suggests that all Leoville fractions sampled here formed within an isotopically similar reservoir in the protoplanetary disk. \\

We have several side notes to this conclusion: 1) Although we have shown that chondrules and matrix could have formed within the same reservoir, the chondrules are not complementary to the matrix, since the matrix is isotopically chondritic, similar to bulk CV rather than super-chondritic; 2) Without analyzing the Zn isotope compositions of chondrules and matrices in other chondrite classes, it is unknown whether the isotopic relationships shown here are unique to the CV chondrite reservoir or represent more generic relationships  \citep{Connelly2018}. It is, therefore, necessary to conduct a systematic study into the Zn isotope compositions of other chondrites. Our new Zn isotope method can now accommodate the analyses of singularly small samples such as CM, CO and EC chondrules ($<$300 $\upmu$m diameter); 3) Elemental complementarity scenarios do not account for more complex chondrule forming scenarios that include multiple melting events, thermal processing, melt-gas interaction and recondensation (Fig. \ref{fig:Scenarios}).

\section{Conclusions}
We report on an adopted Zn isotope analytical protocol to accommodate the measurement of small Zn aliquots (e.g., 2-5 ng of Zn). The newly developed method is successfully tested on various terrestrial standards and samples, which show that accurate and reproducible Zn isotope analyses are dependent on the following size-limiting factors: 1) A low level of contamination of matrix elements in the Zn-cut, 2) low level contamination from resin-derived organics and 3) high signal/noise levels. We further present as a proof of concept, Zn isotope analyses on small fractions (e.g., chondrule cores, igneous rims and matrix) from the relatively unaltered Leoville CV3.1 chondrite. First, we demonstrate that Zn isotope variations from Leoville samples reflect primary signatures related to nebular heterogeneity and/or chondrule forming processes, rather than aqueous alteration. These variations include tightly clustered negative $\updelta$$^{66}$Zn values for the chondrule cores ($\updelta$$^{66}$Zn = --0.43$\pm$0.14 $\permil$), more variable and positive values for the igneous rims ($\updelta$$^{66}$Zn = --0.01$\pm$0.30 $\permil$) and chondritic values for the matrix ($\updelta$$^{66}$Zn = 0.19$\pm$0.14 $\permil$). We further show that combined with the elemental composition and petrology, the Zn isotope analyses of the Leoville fractions point towards the following chondrule forming model as a mechanism to fractionate Zn isotopes: The sulfides containing isotopically heavy Zn must be vaporized during chondrule core formation, thereby lowering the $\updelta$$^{66}$Zn values of the chondrules. Concurrently, chondritic dust is thermally processed alongside the chondrules and subsequently accretes to the chondrule cores as volatile depleted forsterite-bearing grains, which represent the precursors to the igneous rims. These grains react with the complementary volatile-rich gas, which progressively enriches the igneous rims in heavy Zn isotopes as well as moderately volatile elements. This results in a near-chondritic elemental composition of the igenous rims, as well as forsterite relict grains overgrown by low-Ca pyroxene co-existing with troilite/metal assemblages, in agreement with our data. We suggest that all CV components are not complementary to each other in the classical sense proposed by \citep{Hezel2018}, but could have formed from the same chemical and isotopic reservoir. We note that further Zn isotope investigations of other chondrite classes are necessary to distinguish between a generic scenario of isotope fractionation between chondrules and matrix or if these signatures are specific to CV chondrites. Our newly developed Zn isotope analytical protocol will be able to forward this research for small samples such as CM, CO and EC chondrules, and be applied to a large set of small rock samples (f.e., mineral separates and specimens from sample return missions). 

\section*{Acknowledgements}
This project has received funding from the European Union’s Horizon 2020 research and innovation programme under the Marie Sk\l{}odowska-Curie Grant Agreement No 786081. F.M. acknowledges funding from the European Research Council under the H2020 framework program/ERC grant agreement no. 637503 (Pristine) and financial support of the UnivEarthS Labex program at Sorbonne Paris Cit\'{e} (ANR-10-LABX- 0023 and ANR-11-IDEX-0005-02). Parts of this work were supported by IPGP multidisciplinary PARI program, and by Region \^{I}le-de-France SESAME Grant no. 12015908. We thank Jian Huang and two anonymous reviewers for their constructive comments on this manuscript. We thank the Natural History Museum of Denmark for the generous loan of a thick section from Leoville.

\section*{References}
\bibliographystyle{elsarticle-harv} 
\bibliography{Zntechnical2}

\newpage

\begin{sidewaystable}[]
	\centering
	\footnotesize
	\caption{Zn contributions from acids and procedures used during sample digestion and purification. Final blank contributions (right column) reflect the most conservative values from repeat analyses of blanks.}
	\begin{tabular}{rrrllc}
		\multirow{2}{*}{\textbf{Acid}}  & \multirow{2}{*}{\textbf{Volume (ml)}} & \multirow{2}{*}{\textbf{Zn (ng)}} & \multirow{2}{*}{\textbf{RSD \%}} & \multirow{2}{*}{\textbf{Use}}   & \textbf{Blank contribution} \\
		&&&&& \textbf{Zn (ng)}\\
		\toprule
		1.5M HBr (1) & \multicolumn{1}{c}{2.0} & \multicolumn{1}{c}{n.d.} &       & 1.6 ml in two-step  & \textit{\textbf{$<$0.010}} \\
		1.5M HBr (2) & \multicolumn{1}{c}{2.0} & \multicolumn{1}{c}{0.056} & 7.80  &  anion column     &  \\
		\midrule
		6M HCl (1) & \multicolumn{1}{c}{2.0} & \multicolumn{1}{c}{0.041} & 78.24 & 0.5 ml in sample digestion & \textit{\textbf{$<$0.005}} \\
		6M HCl (2) & \multicolumn{1}{c}{2.0} & \multicolumn{1}{c}{n.d.} &       &       &  \\
		\midrule
		0.5M HNO$_{3}$ (1) & \multicolumn{1}{c}{4.0} & \multicolumn{1}{c}{n.d.} &       & 1.6 ml in two-step  & \textit{\textbf{n.d.}} \\
		0.5M HNO$_{3}$ (2) & \multicolumn{1}{c}{4.0} & \multicolumn{1}{c}{n.d.} &       &    anion column   &  \\
		\midrule
		7M HNO$_{3}$ (1) & \multicolumn{1}{c}{2.0} & \multicolumn{1}{c}{0.163} & 8.98  & 1 ml during digestion & \textit{\textbf{$<$0.08}} \\
		7M HNO$_{3}$ (2) & \multicolumn{1}{c}{2.0} & \multicolumn{1}{c}{0.065} & 19.35 &       &  \\
		\midrule
		27M HF (1) & \multicolumn{1}{c}{2.0} & \multicolumn{1}{c}{0.082} & 7.48  & 100 $\upmu$l during digestion & \textit{\textbf{$<$0.004}} \\
		27M HF (2) & \multicolumn{1}{c}{2.0} & \multicolumn{1}{c}{0.018} & 38.54 &       &  \\
		\midrule
		7M HNO$_{3}$ + 27M HF (1) & \multicolumn{1}{c}{1.0} & \multicolumn{1}{c}{0.051} & 6.19  & Digestion of sample in   & \textit{\textbf{$<$0.574}} \\
		7M HNO$_{3}$ + 27M HF (2) & \multicolumn{1}{c}{1.0} & \multicolumn{1}{c}{0.574} & 6.94  &  1 ml HNO$_{3}$/HF in      &  \\
		7M HNO$_{3}$ + 27M HF (3) & \multicolumn{1}{c}{1.0} & \multicolumn{1}{c}{0.202} & 11.35 &  Parr bombs at 210$^{\circ}$C     &  \\
		7M HNO$_{3}$ + 27M HF (4) & \multicolumn{1}{c}{1.0} & \multicolumn{1}{c}{0.116} & 5.48  &       &  \\
		\midrule
		1.5M HBr + 0.5M HNO$_{3}$ (1) & \multicolumn{1}{c}{1.6} & \multicolumn{1}{c}{0.175} & 26.96 & Total procedural blank  & \textit{\textbf{$<$0.350}} \\
		1.5M HBr + 0.5M HNO$_{3}$ (2) & \multicolumn{1}{c}{1.6} & \multicolumn{1}{c}{0.019} & 112.83 & of column chemistry      &  \\
		1.5M HBr + 0.5M HNO$_{3}$ (3) & \multicolumn{1}{c}{1.6} & \multicolumn{1}{c}{0.097} & 16.22 &       &  \\
	1.5M HBr + 0.5M HNO$_{3}$ (4) & \multicolumn{1}{c}{1.6} & \multicolumn{1}{c}{n.d.} &       &       &  \\
	1.5M HBr + 0.5M HNO$_{3}$ (5) & \multicolumn{1}{c}{1.6} & \multicolumn{1}{c}{n.d.} &       &       &  \\
		\bottomrule
		&       &       &       & \textbf{Total Zn blank (with Parr bombs)} & \textit{\textbf{$<$0.924}} \\
		&       &       &       & \textbf{Total Zn blank (no Parr bombs)} & \textit{\textbf{$<$0.350}} \\
	\end{tabular}%
	\label{tab:blanks}%
\end{sidewaystable}%

\begin{longtable}{rcccc}
	\caption{Zn isotope data of terrestrial standards with given amounts of total Zn over the column and amounts of Zn used during individual analyses by MC-ICPMS. Literature data are from \citet{Moynier2017,Inglis2017,Pringle2017}.}\\
\toprule
	\multirow{2}{*}{Standard} & Total amount & Amount of Zn & \multirow{2}{*}{$\updelta$$^{66}$Zn ($\permil$)} & \multirow{2}{*}{$\updelta$$^{68}$Zn ($\permil$)} \\
			& of Zn (ng) & individual analyses (ng) & & \\
	\midrule
	\endfirsthead
	\multicolumn{5}{c}%
	{\tablename\ \thetable\ -- \textit{Continued from previous page}} \\
\midrule
	\multirow{2}{*}{Standard} & Total amount & Amount of Zn & \multirow{2}{*}{$\updelta$$^{66}$Zn ($\permil$)} & \multirow{2}{*}{$\updelta$$^{68}$Zn ($\permil$)} \\
			& of Zn (ng) & individual analyses (ng) & & \\
	\midrule
	\endhead
	\hline \multicolumn{5}{r}{\textit{Continued on next page}} \\
	\endfoot
	\bottomrule
	\endlastfoot
			BHVO-2 & 600   & 5     & 0.39  & 0.69 \\
	
	&       & 5     & 0.35  & 0.59 \\
	&       & 5     & 0.34  & 0.58 \\
	&       & 5     & 0.31  & 0.67 \\
	&       & 5     & 0.27  & 0.71 \\
	&       & 5     & 0.32  & 0.65 \\
	&	&5& 0.35 & 0.83\\
	&&5&0.41&0.75\\
	&&5&0.34&0.70\\
	\textit{\textbf{Average}} &       &       & \textit{\textbf{0.34}} & \textit{\textbf{0.69}} \\
	\textit{2SD} &       &       & \textit{0.08} & \textit{0.16} \\
	&       & 2.5 & 0.31  & 0.32 \\
	&       & 2.5 & 0.27  & 0.30 \\
	&       & 2.5 & 0.26  & 0.28 \\
	&       & 2.5 & 0.24  & 0.27 \\
	&       & 2.5 & 0.28  & 0.31 \\
	&       & 2.5 & 0.31  & 0.37 \\
	&       & 2.5 & 0.26  & 0.46 \\
	&       & 2.5 & 0.27  & 0.41 \\
	&       & 2.5 & 0.26  & 0.33 \\
	\textit{\textbf{Average}} &       &       & \textit{\textbf{0.27}} & \textit{\textbf{0.34}} \\
	\textit{2SD} &       &       & \textit{0.05} & \textit{0.13} \\
	\textit{\textbf{Literature}} &       &       & \textit{\textbf{0.28}} &  \\
	\textit{2SD} &       &       & \textit{0.09} &  \\
	\midrule
	&       &       &       &  \\
	BCR-2-1 & 5     & 5     & 0.06  & 0.18 \\
	BCR-2-2 & 5     & 5     & 0.15  & 0.48 \\
	BCR-2-3 & 5     & 5     & 0.14  & 0.34 \\
	BCR-2-4 & 10    & 5     & 0.22  & 0.48 \\
	&       & 5     & 0.21  & 0.46 \\
	BCR-2-5 & 5     & 5     & 0.20  & 0.12 \\
	\textit{\textbf{Average}} &       &       & \textit{\textbf{0.16}} & \textit{\textbf{0.34}} \\
	\textit{2SD} &       &       & \textit{0.12} & \textit{0.32} \\
	\textit{\textbf{Literature}} &       &       & \textit{\textbf{0.25}} &  \\
	\textit{2SD} &       &       & \textit{0.08} &  \\
	\midrule
	&       &       &       &  \\
	AGV2-1 & 5     & 5     & 0.32  & 0.59 \\
	AGV2-2 & 5     & 5     & 0.25  & 0.21 \\
	AGV2-3 & 10    & 5     & 0.15  & 0.55 \\
	& 10    & 5     & 0.11  & 0.50 \\
	AGV2-4 & 5     & 5     & 0.21  & 1.05 \\
	\textit{\textit{Average}} &       &       & \textit{\textbf{0.21}} & \textit{\textbf{0.58}} \\
	\textit{2SD} &       &       & \textit{0.16} & \textit{0.60} \\
	\textit{\textbf{Literature }} &       &       & \textit{\textbf{0.29}} &  \\
	\textit{2SD} &       &       & \textit{0.06} &  \\
	\midrule
	&       &       &       &  \\
	CP1-1 & 5     & 5     & --0.07 & --0.19 \\
	CP1-2 & 10    & 5     & --0.06 & --0.20 \\
	&       & 5     & --0.08 & --0.06 \\
	CP1-3 & 5     & 5     & --0.05 & --0.16 \\
	CP1-4 & 5     & 5     & 0.01 & --0.18 \\
	CP1-5 & 10    & 5     & --0.10 & --0.24 \\
	&       & 5     & -0.15 & -0.14 \\
	\textit{\textbf{Average}} &       &       & \textit{\textbf{--0.07}} & \textit{\textbf{--0.17}} \\
	\textit{2SD} &       &       & \textit{0.10} & \textit{0.11} \\
	\textit{\textbf{Literature}} &       &       & \textit{\textbf{0.00}} &  \\
	\textit{2SD} &       &       & \textit{0.08} &  \\
	\midrule
	&       &       &       &  \\
	CV chondrite & 600   & 2.5   & 0.33  & 1.36 \\
	&       & 2.5   & 0.36  & 1.59 \\
	&       & 2.5   & 0.17  & 0.17 \\
	&       & 2.5   & 0.14  & 0.06 \\
	&       & 2.5   & 0.39  & 0.55 \\
	&       & 2.5   & 0.25  & 0.21 \\
	&       & 2.5   & 0.21  & 0.02 \\
	&       & 2.5   & 0.13  & 0.06 \\
	&       & 2.5   & 0.27  & 0.42 \\
	&       & 2.5   & 0.48  & 0.65 \\
	&       & 2.5   & 0.21  & 0.31 \\
	&       & 2.5   & 0.26  & 0.44 \\
	\textit{\textbf{Average}} &       &       & \textit{\textbf{0.27}} & \textit{\textbf{0.49}} \\
	\textit{2SD} &       &       & \textit{0.19} & \textit{1.17} \\
	\textit{\textbf{Literature}} &       &       &  \textit{\textbf{0.24}}     &  \textit{\textbf{0.44}}\\
	\textit{2SD} &       &       &   \textit{0.12}    &  \textit{0.21}\\
	\midrule
	JMC Lyon (processed) & 1   & 1   & 0.06  &  \\
		 &    & 1   & --0.08  &  \\
		 &    & 1   & --0.10  &  \\
		 &    & 1   & 0.00  &  \\
		 \textit{\textbf{Average}} &       &       & \textit{\textbf{--0.03}} &  \\
		 \textit{2SD} &       &       & \textit{0.15} &  \\
	\label{tab:standardsZn}
\end{longtable}

\begin{sidewaystable}[htbp]
	\centering
	\caption{Major and minor element ratios of Leoville chondrule fractions. Analyses are carried out by Agilent 7900 ICP-Q-MS. The error is a percentage calculated from the percentage offset between terrestrial standards and literature values. The offset is taken from BCR-2, since AGV2 likely experienced some element loss to the Zn-cut. }
	\begin{tabular}{rccccccccccc}
		& \multicolumn{5}{c}{\textbf{Standards}}         & \multicolumn{6}{c}{\textbf{Cores}} \\
		\toprule
		& \multicolumn{1}{l}{AGV2} & \multicolumn{1}{l}{AGV2$_{lit}$} & \multicolumn{1}{l}{BCR-2} & \multicolumn{1}{l}{BCR-2$_{lit}$} & \multicolumn{1}{l}{Error (\%)} & 1     & 2     & 3     & 5     & 6     & \multicolumn{1}{l}{average} \\
		\midrule
		&       &       &       &       &       &       &       &       &       &       &  \\
		Na/Mg & 2.97  & 2.88  & 1.05  & 1.08  & 2.79  & 0.038 & 0.008 & 0.030 & 0.024 & 0.021 & 0.024 \\
		Al/Mg & 8.66  & 8.29  & 3.28  & 3.31  & 0.75  & 0.35  & 0.10  & 0.22  & 0.10  & 0.10  & 0.18 \\
		Ca/Mg & 3.76  & 3.44  & 2.42  & 2.36  & 2.61  & 0.34  & 0.08  & 0.22  & 0.12  & 0.10  & 0.17 \\
		Cr/Mg & 0.0023 & 0.0016 & 0.00085 & 0.00083 & 1.46  & 0.020 & 0.016 & 0.012 & 0.025 & 0.017 & 0.018 \\
		Ti/Mg & 0.47  & 0.58  & 0.644 & 0.625 & 2.89  & 0.0139 & 0.0039 & 0.0115 & 0.0041 & 0.0056 & 0.0078 \\
		Mn/Mg & 0.07  & 0.07  & 0.067 & 0.070 & 5.48  & 0.0050 & 0.0036 & 0.0021 & 0.0035 & 0.0022 & 0.0033 \\
		Fe/Mg & 4.10  & 4.33  & 4.16  & 4.47  & 6.37  & 0.41  & 0.61  & 0.11  & 1.44  & 0.28  & 0.57 \\
		Ni/Mg & 0.0021 & 0.0018 & 0.0006 & 0.0005 & 27.42 & 0.010 & 0.028 & 0.002 & 0.095 & 0.027 & 0.032 \\
		Co/Mg & 0.0016 & 0.0015 & 0.0019 & 0.0017 & 8.29 & 0.0063 & 0.0027 & 0.0320 & 0.0139 & 0.0072 & 0.0124 \\
		\bottomrule
		&       &       &       &       &       &       &       &       &       &       &  \\
		& \multicolumn{6}{c}{\textbf{Rims}}                      & \multicolumn{3}{c}{\textbf{Matrix}} &       &  \\
		\toprule
		& 1     & 2     & 3     & 5     & 6     & \multicolumn{1}{l}{average} & 1     & 2     & \multicolumn{1}{l}{average} &       &  \\
		\midrule
		&       &       &       &       &       &       &       &       &       &       &  \\
		Na/Mg & 0.022 & 0.010 & 0.030 & 0.022 & 0.022 & 0.021 & 0.034 & 0.044 & 0.039 &       &  \\
		Al/Mg & 0.14  & 0.10  & 0.10  & 0.08  & 0.12  & 0.108 & 0.19  & 0.10  & 0.15  &       &  \\
		Ca/Mg & 0.15  & 0.09  & 0.12  & 0.09  & 0.11  & 0.113 & 0.21  & 0.10  & 0.15  &       &  \\
		Cr/Mg & 0.039 & 0.025 & 0.036 & 0.033 & 0.035 & 0.033 & 0.042 & 0.030 & 0.036 &       &  \\
		Ti/Mg & 0.0077 & 0.0055 & 0.0061 & 0.0047 & 0.0053 & 0.0059 & 0.0073 & 0.0037 & 0.0055 &       &  \\
		Mn/Mg & 0.0085 & 0.0192 & 0.0037 & 0.0156 & 0.0048 & 0.0125 & 0.0429 & 0.0192 & 0.0311 &       &  \\
		Fe/Mg & 2.31  & 1.47  & 1.17  & 2.16  & 1.14  & 1.65  & 3.19  & 3.35  & 3.27  &       &  \\
		Ni/Mg & 0.144 & 0.054 & 0.090 & 0.115 & 0.126 & 0.106 & 0.101 & 0.131 & 0.116 &       &  \\
		Co/Mg & 0.0124 & 0.0036 & 0.0297 & 0.0074 & 0.0124 & 0.0131 & 0.0064 & 0.0069 & 0.0067 && \\
		\bottomrule
	\end{tabular}%
	\label{tab:majorel}%
\end{sidewaystable}%

\begin{table}[htbp]
	\centering
	\caption{Zn isotope data for Leoville chondrule cores, igneous rims and surrounding matrices, with the abundance of Zn given for each sample. Note that three chondrule cores only had 1 ng of Zn and were analyzed in a block of 15 cycles on the MC-ICPMS, rather than the 30 cycles used for the other samples. Standards measured with 15 cycles yield identical $\updelta$$^{66}$Zn values within 0.02 $\permil$.}
	\begin{tabular}{lrcccc}
		\toprule
		&   & Zn (ng)  & Cycles  & \textbf{$\updelta$$^{66}$Zn ($\permil$)} & \textbf{$\updelta$$^{68}$Zn ($\permil$)} \\
		\midrule
		\multicolumn{1}{l}{Matrix} & ch1  & 5 & 30 & 0.24$\pm$0.12& 0.51$\pm$0.32\\
		& ch2 & 5 & 30 & 0.14$\pm$0.12& 0.59$\pm$0.32\\
		& \textit{\textbf{Mean}} &&& \textit{\textbf{0.19}}$\pm$\textit{\textbf{0.14}}& \textit{\textbf{0.55}}$\pm$\textit{\textbf{0.12}}\\
		&       &    &&   &  \\
		\multicolumn{1}{l}{Rims} & ch1 & 5 & 30 & --0.14$\pm$0.12& 0.54$\pm$1.17\\
		& ch2 & 2 & 30 & --0.11$\pm$0.19 & 1.89$\pm$0.32\\
		& ch3  & 2 & 30& --0.04$\pm$0.19 & 1.08$\pm$1.17\\
		& ch5  & 5 & 30& 0.23$\pm$0.12 & 0.45$\pm$1.17\\
		& ch6  & 3& 30 & 0.01$\pm$0.19 & 0.76$\pm$1.17\\
		& \textit{\textbf{Mean}} &&& \textit{\textbf{--0.01}}$\pm$\textit{\textbf{0.30}} & \textit{\textbf{0.94}}$\pm$\textit{\textbf{1.16}} \\
		&       &     &&  &  \\
		\multicolumn{1}{l}{Cores} & ch1 & 1 & 15 & --0.41$\pm$0.19 & 0.14$\pm$1.17\\
		& ch2 & 1 & 15 & --0.41$\pm$0.19 &0.10$\pm$1.17\\
		& ch3 & 2 & 30 & --0.54$\pm$0.19 &0.40$\pm$1.17\\
		& ch5 & 1 & 15 & --0.36$\pm$0.19 & --0.15$\pm$1.17\\
		& ch6 & 3 & 30 & --0.41$\pm$0.12 &0.04$\pm$0.32\\
		& \textit{\textbf{Mean}} &&& \textit{\textbf{--0.43}} $\pm$ \textit{\textbf{0.14}} & \textit{\textbf{0.10}}$\pm$\textit{\textbf{0.40}}\\
		\bottomrule
	\end{tabular}%
	\label{tab:LeovilleZn}%
\end{table}%

\begin{table}[htbp]
	\centering
	\caption{Matrix Fe/Mg and Mn/Mg ratios of various CV chondrites analyzed by different authors and using different analytical techniques. DBA = Defocused or broad beam analyses, WA = Wet chemical analyses, INAA = Instrumental neutron activation analysis.}
	\begin{tabular}{lccll}
			\toprule
		\textbf{Sample} & \multicolumn{1}{l}{\textbf{Fe/Mg}} & \multicolumn{1}{l}{\textbf{Mn/Mg}} & \textbf{Authors} & \textbf{Technique} \\
		\midrule
		Leoville & 2.11  & 0.015 & \citet{Kracher1985} & DBA \\
		Leoville & 1.96  & 0.016 & \citet{McSween1977} & DBA \\
		Leoville & 3.27 & 0.031 & This study & WA\\
		Vigarano & 2.46  & 0.011 & \citet{McSween1977} & DBA \\
		Vigarano & 3.23  & 0.019 & \citet{Zolensky1993} & DBA \\
		Vigarano & 2.71  & 0.022 & \citet{Klerner2001} & DBA \\
		Efremovka & 2.63  & 0.015 & \citet{McSween1977} & DBA \\
		Efremovka & 2.53  & 0.022 & \citet{Hezel2010} & DBA \\
		Allende & 1.79  & 0.013 & \citet{Clarke1971} & WA \\
		Allende & 2.03  & 0.013 & \citet{McSween1977} & DBA \\
		Allende & 2.36  & 0.014 & \citet{Rubin1984} & DBA \\
		Allende & 1.57 & 0.011 & \citet{Rubin1987} & INAA\\
		\bottomrule
	\end{tabular}%
	\label{tab:Matrix}%
\end{table}%

\newpage

\end{document}